\documentclass[12pt,fleqn]{article}
\usepackage{amsfonts}
\usepackage{amssymb}
\usepackage{amsmath}
\usepackage{bm}
\usepackage{geometry}
\usepackage{setspace}
\usepackage{caption}
\usepackage[multiple]{footmisc}

\usepackage{graphicx}   % to manage images in Latex/Overleaf
\usepackage{textcomp}   % to use currency symbols (\textcent)

\begin{document}

\title{Detecting Edgeworth Cycles\thanks{%
We thank Xiaohong Chen for advice on time-series methods, Cuicui Chen for
detailed discussions, Robert Clark, Daniel Ershov, Simon Martin, and Felix
Montag for advice on the German data, Kevin Sch\"{a}fer and the Argus Media
group for kindly providing us with the German wholesale prices, and our team
of research assistants at Yale University (Yue Qi, Alan Chiang, Alexis Teh,
Clara Penteado, Janie Wu, Bruno Moscarini, Eileen Yang, and Jordan Mazza)
for excellent work. We also thank Roxana Mihet and seminar/conference
participants at the D\"{u}sseldorf Institute for Competition Economics
(DICE), Brown University, Cornell University, Yale University, IIOC 2022,
and ES-NASM 2022 for comments. This work was supported by the Swiss National
Science Foundation (SNF), under project ID \textquotedblleft New methods for
asset pricing with frictions.\textquotedblright }}
\author{Timothy Holt\thanks{%
Institute of Computing, Universit\`{a} della Svizzera italiana. E-mail:
timothy.holt@usi.ch.} \and Mitsuru Igami\thanks{%
Department of Economics, Yale University. E-mail: mitsuru.igami@yale.edu.}
\and Simon Scheidegger\thanks{%
Department of Economics, HEC Lausanne. E-mail: simon.scheidegger@unil.ch.}}
\date{July 5, 2022}
\maketitle

\begin{abstract}
We develop and test algorithms to detect \textquotedblleft Edgeworth
cycles,\textquotedblright \ which are asymmetric price movements that have
caused antitrust concerns in many countries. We formalize four existing
methods and propose six new methods based on spectral analysis and machine
learning. We evaluate their accuracy in station-level gasoline-price data
from Western Australia, New South Wales, and Germany. Most methods achieve
high accuracy in the first two, but only a few can detect the nuanced cycles
in the third. Results suggest whether researchers find a positive or
negative statistical relationship between cycles and markups, and hence
their implications for competition policy, crucially depends on the choice
of methods. We conclude with a set of practical recommendations.

\bigskip

\noindent \textit{Keywords}: Deep neural networks, Edgeworth cycles, Fuel
prices, Machine learning, Markups, Nonparametric methods, Spectral analysis

\bigskip

\noindent \textit{JEL classifications}: C45 (Neural Networks and Related
Topics), C55 (Large Data Sets: Modeling and Analysis), L13 (Oligopoly and
Other Imperfect Markets), L41 (Monopolization, Horizontal Anticompetitive
Practices).
\end{abstract}

\section{Introduction}

Retail gasoline prices are known to follow cyclical patterns in many
countries (e.g., Byrne and de Roos 2019). The patterns persist even after
controlling for wholesale and crude-oil prices. Because these cycles are so
regular and conspicuous, and because price increases tend to be larger than
decreases, observers suspect anti-competitive business practices. The
occasional discovery of price-fixing cases supports this view (e.g., Clark
and Houde 2014, Foros and Steen 2013, Wang 2008).\footnote{%
Recent studies on algorithmic collusion suggest interactions between
self-learning algorithms could lead to collusive equilibria with such cycles
(Klein 2021); the use of \textquotedblleft repricing
algorithms\textquotedblright \ by many sellers on Amazon has made these
phenomena prevalent in e-commerce as well (Musolff 2021).}

These asymmetric movements are called Edgeworth cycles and have been studied
extensively.\footnote{%
Maskin and Tirole (1988) coined the term after Edgeworth's (1925)
hypothetical example. It became a popular topic for empirical research since
Castanias and Johnson (1993). We explain its theoretical background in
section 2.} In particular, scholars and antitrust practitioners have
investigated whether the presence of cycles is associated with higher prices
and markups. Deltas (2008), Clark and Houde (2014), and Byrne (2019) find
that asymmetry is correlated with higher margins, price-fixing collusion,
and concentrated market structure, respectively. However, Lewis (2009),
Zimmerman et al. (2013), and Noel (2015) show prices and margins are \textit{%
lower} in markets with asymmetric price cycles. Given the diversity of
countries and regions in these studies (Australia, Canada, the US, and
several countries in Europe), the cycle-competition relationship could be
intrinsically heterogeneous across markets.

But another, perhaps more fundamental, problem is measurement: the lack of a
formal definition or a reliable method to detect cycles in large datasets.
Because theory provides only a loose characterization of Edgeworth cycles,
empirical researchers have to rely on visual inspections and summary
statistics based on a single quantifiable characteristic: asymmetry.
Meanwhile, the phenomena's most basic property, cyclicality, is almost
completely absent from the existing operational definitions. Even though
asymmetry may be the most salient feature of---and hence a necessary
condition for---Edgeworth cycles, it is not a sufficient condition.
Empirical findings are only as good as the measures they employ; the
incompleteness of detection methods could affect the reliability of
\textquotedblleft facts\textquotedblright \ about competition and price
cycles. Now that the governments of many countries and regions are making
large-scale price data publicly available,\footnote{%
The governments of Australia, Germany, and other countries have made
detailed price data publicly available to inform consumers and encourage
further scrutiny. The Australian Consumer and Competition Commission has a
team dedicated to monitoring gasoline prices and regularly publishes
reports. See
https://www.accc.gov.au/consumers/petrol-diesel-lpg/about-fuel-prices. The
Bundeskartellamt does the same in Germany.} developing a scalable detection
method represents an important practical challenge for economists and
policymakers.

This paper proposes a systematic approach to detecting Edgeworth cycles. We
formalize four existing methods as simple parametric models: (1) the
\textquotedblleft positive runs vs. negative runs\textquotedblright \ method
of Castanias and Johnson (1993), (2) the \textquotedblleft mean increase vs.
mean decrease\textquotedblright \ method of Eckert (2002), (3) the
\textquotedblleft negative median change\textquotedblright \ method of Lewis
(2009), and (4) the \textquotedblleft many big price
increases\textquotedblright \ method of Byrne and de Roos (2019). We then
propose six new methods based on spectral analysis and
nonparametric/machine-learning techniques: (5) Fourier transform, (6) the
Lomb-Scargle periodogram, (7) cubic splines, (8) long short-term memory
(LSTM), (9) an \textquotedblleft ensemble\textquotedblright \ (aggregation)
of Methods 1--7 within a random-forests framework, and (10) an ensemble of
Methods 1--8 within an extended LSTM.\footnote{%
Section 4 formally introduces all methods.}

To evaluate the performance of each method, we collect data on retail and
wholesale gasoline prices in two regions of Australia, Western Australia
(WA) and New South Wales (NSW), as well as the entirety of Germany. These
datasets cover the universe of gasoline stations in these regions/countries,
record each station's retail price at a daily (or higher) frequency, and are
made publicly available by legal mandates.\footnote{%
See Byrne, Nah, and Xue (2018) for a guide to the Australian data. Haucap,
Heimeshoff, and Siekmann (2017), Martin (2018), and Assad, Clark, Ershov,
and Xu (2021), among others, study the German data.} Given the lack of a
clear theoretical definition, we construct a benchmark \textquotedblleft
ground truth\textquotedblright \ based on human recognition of price cycles
as follows. We reorganize the raw data as panel data of the daily margins (=
retail minus wholesale prices) of gasoline stations and group them into
calendar quarters, so that a station-quarter (i.e., a set of 90 consecutive
days of retail-margin observations for each station) becomes the effective
unit of observation.\footnote{%
We explain our data, the choice of sampling frequency, and the manual
classification procedures in section 3.} We employ eight research assistants
(RAs) to manually classify each station-quarter as either \textquotedblleft
cycling,\textquotedblright \  \textquotedblleft maybe
cycling,\textquotedblright \ or \textquotedblleft not
cycling.\textquotedblright \ We then define a binary indicator variable that
equals 1 if an observation is labeled as \textquotedblleft
cycling\textquotedblright \ by all of the RAs (the majority of observations
are labeled by three RAs), and 0 otherwise, thereby preparing a conservative
target for automatic cycle detection.\footnote{%
Appendix B.3 shows results under an alternative criterion.} Note that we
look only for cyclicality and do not impose asymmetry or other criteria in
the manual-classification stage. The reason is that asymmetry is---unlike
cyclicality---amenable to clear mathematical definitions and can easily be
checked at a later stage. Hence, we prioritize the detection of cyclicality,
thereby alleviating the cognitive burden on RAs.

At this point, one might wonder whether human recognition of cycles is an
appropriate benchmark. We regard it as the best \textit{feasible} option
(the \textquotedblleft second best\textquotedblright ) given the lack of
clear theoretical definitions (the \textquotedblleft first
best\textquotedblright ). Manual classification by a team of RAs represents
a best-effort practice in the literature and provides a relevant---if not
perfect---benchmark in the following sense. First, most existing studies
employ some rule-of-thumb definitions with calibrated thresholds, which are
ultimately based on the researchers' eyeballing and judgment, the details of
which are rarely documented. We make such procedures more explicit,
systematic, and transparent, so that the overall scheme becomes more
reproducible. Second, human recognition is central to the prominence of
Edgeworth cycles as an antitrust topic. Despite the lack of universal
definitions, the phenomena have become a perennial policy issue in many
countries precisely because consumers and politicians can easily recognize
cyclical patterns when they see them. In this regard, human recognition 
\textit{is} the \textquotedblleft ground truth\textquotedblright \ that
eventually determines the phenomena's relevance to public policy. We
interpret our RAs' responses as a proxy for the general public's responses
to various patterns in gasoline prices.

\begin{figure}[htb!!!!]\centering%
\caption{Examples of Cycling and Non-cycling Station-Quarter Observations}%
\includegraphics[width=1.00\textwidth]{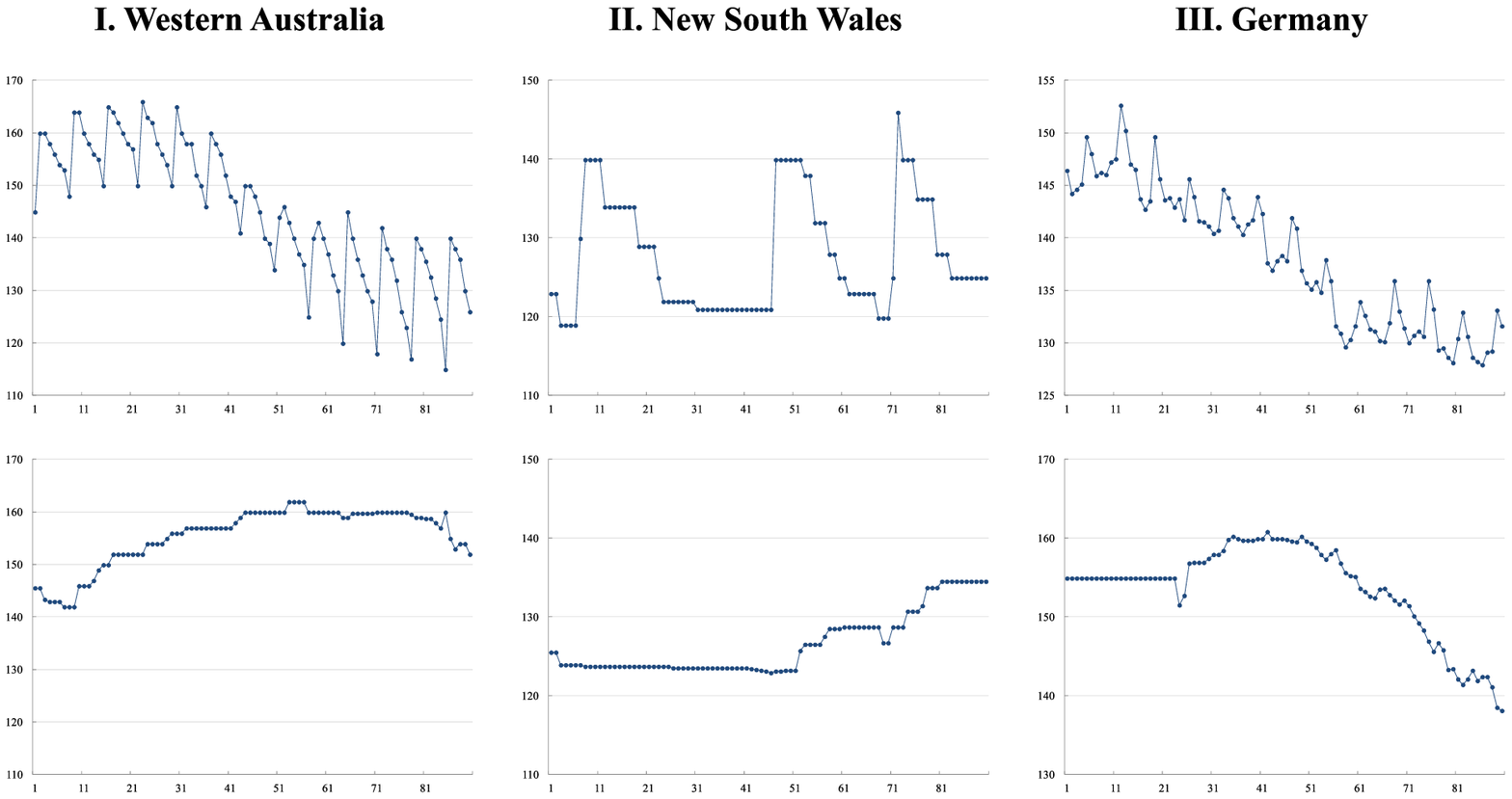}
\caption*{\footnotesize {%
\textit{Note}: The top panels and the bottom panels show examples of daily
retail-price series in \textquotedblleft cycling\textquotedblright \ and
\textquotedblleft non-cycling\textquotedblright \ station-quarter
observations, respectively, for illustration purposes. The vertical axes
measure retail gasoline prices in the Australian cent (left, middle) and the
Euro cent (right), respectively. The horizontal axes represent calendar
days. Note our main analysis uses retail margins (= retail minus wholesale
prices) instead, thereby controlling for costs.}}%
\label{Figure - Examples}
\end{figure}

We report three sets of results. First, when applied to the two Australian
datasets, most of the methods---both existing and new---achieve high
accuracy levels near or above 90\% and 80\%, respectively, because price
cycles are clearly asymmetric and exhibit regular periodicity (hence, are
easy to detect) in these regions. By contrast, German cycles are more subtle
and diverse, defying many methods. All existing methods except Method 4 fail
to detect cycles, even though as much as 40\% of the sample is unanimously
labeled as \textquotedblleft cycling\textquotedblright \ by three RAs (see
Figure \ref{Figure - Examples} for examples). This failure is not an
artifact of sample selection or human error because our interview with a
German industry expert suggests Edgeworth cycles are known to exist. They
are (in fact) called the \textquotedblleft price parachute\textquotedblright
\ (or \textit{Preis Fallschirm}) phenomena, and are considered to be part of
common pricing strategies among practitioners. Bundeskartellamt (2011) also
confirms the existence of both weekly and daily cycles.\footnote{%
See sections 3.3 and 7.2 for further details on the German data.} Methods
7--10 attain 71\%--80\% accuracy even in this challenging environment.

Second, we assess the cost effectiveness of each method by using only $0.1\%$%
, $1\%$, $5\%$, $10\%$, $\cdots $, $80\%$ of our manually labeled subsamples
as \textquotedblleft training\textquotedblright \ data. Results suggest
simpler models (Methods 1--7) are extremely \textquotedblleft
cheap\textquotedblright \ to train, as they quickly approach their
respective maximal accuracy with only a dozen observations. The
nonparametric models (Methods 8--10) need more data to achieve near-maximal
performance, but their data requirement is sufficiently small for practical
purposes. Only a few hundred observations prove sufficient for even the most
complex model (Method 10). The economic cost of manually classifying a few
hundred observations is in the order of tens of RA hours, or a few hundred
US dollars at the current hourly wage of US\$13.50 for undergraduate RA work
at Yale University. Potential cost savings are sizable, as manually labeling
the entire German dataset in 2014--2020 would require 4,800 RA hours, or
US\$64,800. Thus, our approach is economical and suitable for researchers
and governments with limited resources.

Third, we investigate whether and how gasoline stations' markups are
correlated with the presence of cycles. In WA and NSW, the average margins
in (manually classified) \textquotedblleft cycling\textquotedblright \
station-quarters are statistically significantly higher than in
\textquotedblleft non-cycling\textquotedblright \ ones. The relationship is
reversed in Germany, where the margins in \textquotedblleft
cycling\textquotedblright \ observations are lower than in \textquotedblleft
non-cycling\textquotedblright \ ones. Hence, in general, the presence of
cycles could be either positively or negatively correlated with markups. All
of the automatic detection methods lead to the correct finding (i.e.,
positive correlations) in WA, but some of them fail in NSW. Furthermore,
Methods 1--6 either fail to detect cycles or lead to false conclusions in
Germany (i.e., find statistically significant \textit{positive}
correlations). This finding emerges under both \textquotedblleft cyclicality
only\textquotedblright \ and \textquotedblleft cyclicality with
asymmetry\textquotedblright \ definitions of Edgeworth cycles. Thus, whether
researchers discover\ a positive, negative, or no statistical relationship
between markups and cycles---a piece of highly policy-relevant empirical
evidence---depends on the seemingly innocuous choice of operational
definitions.

\bigskip

The rest of the paper is organized as follows. Sections 2--4 explain the
theoretical background, data, and methods, respectively. Sections 5--7
report our main findings and discuss their economic/policy implications.
Section 8 summarizes our practical recommendations for cycle detection.
Section 9 concludes.

\section{Theoretical Background}

Even though our primary goal is empirical, some conceptual anchoring
clarifies the target of measurement.

\subsection{What Are Edgeworth Cycles?}

Maskin and Tirole (1988) offer the following verbal description:
\textquotedblleft In the Edgeworth cycle story, firms undercut each other
successively to increase their market share (price war phase) until the war
becomes too costly, at which point some firm increases its price. The other
firms then follow suit (relenting phase), after which price cutting begins
again. The market price thus evolves in cycles\textquotedblright \ (pages
571--572). This description and its micro foundation---as a class of Markov
perfect equilibria (MPE) in an alternating-move dynamic duopoly
game---suggest four important characteristics: cyclicality, asymmetry,
stochasticity, and strategicness.

\paragraph{Property 1: Cyclicality.}

The price should exhibit cyclicality, as the terminology suggests. However,
this property is not so obvious in Edgeworth's (1925) original conjecture.
His writing focuses on the indeterminacy of static equilibrium in a
price-setting game between capacity-constrained duopolists. Even though he
mentions a price path that resembles Maskin and Tirole's description as an
example, he uses the word \textquotedblleft cycle\textquotedblright \ only
once. More generally, he conjectures that \textquotedblleft there will be an
indeterminate tract through which the index of value will oscillate, or
rather will vibrate irregularly for an indefinite length of
time\textquotedblright \ (page 118). Thus, Edgeworth's original theory
features not so much cyclicality as \textquotedblleft perpetual
motion\textquotedblright \ (page 121).

Nevertheless, we have chosen to focus on cyclicality in this paper.
Theoretically, Maskin and Tirole's equilibrium strategies (their equation
23) explicitly feature price cycles. Empirically, it is this repetitive
pattern that draws consumers' and politicians' attention; \textquotedblleft
perpetual motion\textquotedblright \ alone would not raise antitrust
concerns.

\paragraph{Property 2: Asymmetry.}

The second characteristic is the asymmetry between relatively few large
price increases and many small price decreases. Edgeworth (1925) does not
emphasize this property either, but it plays an important role in the
Maskin-Tirole formalization and the subsequent empirical literature (see
Methods 1--4 in section 4.1).

\paragraph{Property 3: Stochasticity.}

In Maskin and Tirole's Edgeworth-cycle MPE, big price increases are supposed
to happen stochastically, not deterministically. The reason is that if one
firm always \textquotedblleft relents\textquotedblright \ whenever the low
price is reached, the other firm will always wait and free-ride, which in
turn would make the first firm more cautious about the timing of price
increases. Thus, the frequency of cycles must be stochastic---with varying
lengths of time spent at the low price---in equilibrium.\footnote{%
This theoretical property seems largely overlooked in the empirical
literature, presumably because the first two properties make the phenomena
sufficiently interesting and policy-relevant.} We do not impose stochastic
frequencies as a necessary condition in our empirical procedures, but some
of our methods are designed to accommodate cycles with varying frequencies
(Methods 7 and 8 in section 4.2).

\paragraph{Property 4: Strategicness.}

The cyclical patterns are supposed to emerge from dynamic strategic
interactions between oligopolistic firms. If similar patterns are observed
under monopoly, their underlying mechanism must be different from that of
Edgeworth cycles.\footnote{%
Alternative explanations include consumers with heterogeneous search costs,
intertemporal price discrimination, and \textquotedblleft dynamic
pricing\textquotedblright \ algorithms (broadly defined as any pricing
strategy and its implementation(s) that tries to exploit consumer
heterogeneity and time-varying price-elasticity of demand).} Thus, whether
market structure is monopolistic or oligopolistic is a theoretically
important distinction. Empirically, however, market definition is rarely
clear-cut in practice. Even when a gasoline station is located in a
geographically isolated place, pricing decisions at large chains tend to be
centralized at the city, region, or country level. Market structure at these
aggregate levels is oligopolistic in all of our datasets. Consequently, we
do not impose any geographical boundaries a priori. We simply analyze data
at the individual station level.\footnote{%
This operational decision is not without its own risks. For example, if the
grid of relevant prices were very coarse and two firms take turns to change
prices, we might not be able to observe clear cycles at any specific
station's time-series data even if such cycles exist at the aggregate level.
Fortunately, gasoline prices reside on a relatively fine grid with the
minimum interval of the Australian or Euro cent. Moreover, Maskin and
Tirole's Edgeworth-cycle MPE requires a fine grid with sufficiently small
intervals (denoted by $k$ in their model). Therefore, we believe the risk of
missing aggregate cycles is low.} Our idea is that once the station-level
characterization is successfully completed, one can always compare
cyclicality across stations in the same market (defined geographically or
otherwise) and look for synchronicity---whenever such analysis becomes
necessary.

\subsection{Are Edgeworth Cycles Competitive or Collusive?}

Whether Edgeworth cycles represent collusion is a subtle issue on which we
do not take a stand. Several reasons contribute to its subtlety and our
cautious attitude.

First, the theoretical literature seems agnostic about the distinction
between competitive and collusive behaviors in the current context. On the
one hand, Edgeworth's (1925) narrative lacks any hint of cooperative actions
or intentions. On the other hand, Maskin and Tirole (1988) seem open to
collusive interpretations: \textquotedblleft Several of the results of this
paper underscore the relatively high profits that firms can earn when the
discount factor is near 1. Thus our model can be viewed as a theory of tacit
collusion\textquotedblright \ (page 592). In the more recent literature,
however, the term \textquotedblleft tacit collusion\textquotedblright \ is
usually associated with collusive equilibria in repeated-games models.%
\footnote{%
Tirole and his coauthors exclusively focus on the repeated-games theory when
they summarize the \textquotedblleft economics of tacit
collusion\textquotedblright \ for the European competition authority. See
Ivaldi, Jullien, Rey, Seabright, and Tirole (2003).} The latter rely on the
concepts of monitoring, punishment, and history-dependent strategies as
their underlying mechanism, none of which are prominently featured in
Edgeworth cycles. Thus, even though Maskin and Tirole's own remarks suggest
the possibility of collusive interpretations, we feel inclined to regard
their Edgeworth-cycle MPE as a reflection of competitive interaction between
forward-looking oligopolists.

Second, in terms of antitrust law, explicit communications of a cooperative
nature are the single most important act that constitutes criminal
price-fixing. That is, tacit collusion is not illegal as long as it truly
lacks explicit communication. Notwithstanding this legal distinction, most
of the theoretical literature does not discriminate between tacit and
explicit collusion because the process through which firms reach collusive
agreements is usually not modeled. Hence, an important gap lies between
economic theory and legal enforcement, which complicates the interpretation
of Edgeworth cycles in empirical research.

Third, partly reflecting this unresolved theory-enforcement divide, the
empirical literature has documented many different instances of asymmetric
price cycles, both \textit{with} and \textit{without} legally established
evidence of criminal price-fixing. Accordingly, interpretations of observed
cycles vary across papers on a case-by-case basis. The only common thread
that unites the large empirical literature is the data patterns with clear
cyclicality and asymmetry.

For these reasons, we do not (necessarily) interpret Edgeworth cycles as
evidence of collusion. Consequently, we do not aim or claim to detect
\textquotedblleft collusion.\textquotedblright \ Reliable methods to detect
price cycles would nevertheless be useful for detecting cycle-based
collusion.

\subsection{The Aim of This Paper is to Develop and Compare Data-Analysis
Tools to Capture Cycles}

This paper purposefully avoids theoretical/legal interpretations of the
phenomena as either collusive or competitive. Instead, we concentrate our
efforts on the most basic empirical problem of detecting price cycles. Given
the abundance of asymmetry-based methods in the existing literature, we put
more emphasis on detecting cyclicality. Accordingly, our empirical procedure
starts from the most relaxed characterization of Edgeworth cycles by
constructing a \textquotedblleft ground truth\textquotedblright \ based on
human recognition of cyclicality without any additional criteria. The reason
is not because we disregard the importance of the other properties. Rather,
because Properties 2--4 are amenable to clear mathematical definitions, one
can easily refine the classification of data by imposing asymmetry (and
other formal conditions) \textit{after} a subset of data is identified as
\textquotedblleft cycling.\textquotedblright \ By contrast, the detection of
cyclicality poses a nontrivial challenge and has not been systematically
addressed in the existing literature. The next two sections explain our
approach.

\section{Data and Manual Classification}

Retail-price data are publicly available for the universe of individual
gasoline stations in WA, NSW, and Germany. We combine them with
wholesale-price data, based on the region of each station (Australia) or the
location of the nearest refinery (Germany). We compute station-level daily
profit margins by subtracting the relevant wholesale price from the retail
price,%
\begin{equation}
p_{i,d}\equiv p_{i,d}^{R}-p_{i,d}^{W},
\end{equation}%
where $p_{i,d}^{R}$ and $p_{i,d}^{W}$ are retail and wholesale prices at
station $i$ on day $d$, and simply refer to this markup measure $p_{i,d}$\
as \textquotedblleft price\textquotedblright \ in the following. We organize
these daily prices by calendar quarter, so that station-quarter (i.e., a
sequence of daily prices over 90 days for each station) becomes the unit of
observation for cycle detection.

\subsection{Data Sources and Preparation}

\paragraph{Retail Prices.}

We use three datasets on retail gasoline prices that are publicly available
and of high quality. \textit{FuelWatch} and \textit{FuelCheck} are
legislated retail-fuel-price platforms operated by the state governments of
WA and NSW, respectively. Their websites display real-time information on
petrol prices, and the complete datasets can be downloaded.\footnote{%
Their URLs are https://www.fuelwatch.wa.gov.au and
https://www.fuelcheck.nsw.gov.au.} The Market Transparency Unit for Fuels of
the \textit{Bundeskartellamt} publishes similar data for every German gas
station in minute intervals.\footnote{%
https://www.bundeskartellamt.de/EN/Economicsectors/MineralOil/MTU-Fuels/mtufuels\_node.html%
}

\paragraph{Sampling Frequencies.}

The raw data from WA contain daily retail prices for each station, which is
the most granular level in this region because its law mandates each station
must commit to a fixed price level for 24 hours. By contrast, the stations
in NSW and Germany can change prices at any point in time, which we
aggregate into daily prices by taking either end-of-day values (NSW) or
intra-day averages (Germany). Intra-day changes are relatively rare in NSW,
and hence, end-of-day values are representative of the actual transaction
prices. In Germany, many stations change prices multiple times during the
day, so we sample 24 hourly prices and take their average for each
station-day (see section 3.3 for further details on Germany).

\paragraph{Wholesale Prices.}

The Australian Institute of Petroleum publishes average regional wholesale
prices at https://www.aip.com.au. The Argus Media group's \textit{OMR Oil
Market Report} collects daily regional wholesale prices and offers the
database on a commercial basis.

\subsection{Manual-Classification Procedures}

Whereas most existing studies treat the manual-verification process as an
informal preparatory step (to be embodied by the analyst's eventual choice
of methods and calibration of threshold parameters), we make it as
systematic as possible. Our goal is to develop and compare the performance
of multiple methods, and such \textquotedblleft horse
racing\textquotedblright \ requires a common benchmark.

To establish a \textquotedblleft ground truth\textquotedblright \ based on
human recognition of cycles, we employed a team of eight RAs to manually
classify station-quarter observations.\footnote{%
All of them are graduate or undergraduate students majoring in economics,
mathematics, and statistics at Yale University.} Each station $i$ in quarter 
$t$ is classified as either \textquotedblleft cycling,\textquotedblright \
\textquotedblleft maybe cycling,\textquotedblright \ or \textquotedblleft
not cycling.\textquotedblright \ The total number of manually labeled
observations is 24,569 (WA), 9,693 (NSW), and 35,685 (Germany). The RAs'
total working hours are approximately 260 (WA), 210 (NSW), and 480
(Germany). The manual labeling of the datasets proceeded in three stages.

\paragraph{WA.}

First, we labeled all station-quarters in the WA data with two RAs as a
pilot project between July 2019 and June 2020. The first RA (a PhD student
in economics) laid the ground work with approximately half of the WA data in
close communication with one of the coauthors (Igami). The second RA (a
senior undergraduate student majoring in economics) followed these examples
to label the rest. Then, the first RA carefully double-checked all labels to
maintain consistency. As a result, each station-quarter $\left( i,t\right) $
in WA has one label based on the consensus of the two RAs.

\paragraph{NSW.}

Second, the NSW dataset is smaller but contains more ambiguous cases. Hence,
we took a more organized/computerized approach by building a cloud-based
computational platform to streamline the labeling process. The same coauthor
manually labeled a random sample of 100 station-quarters in December 2020,
which is used for generating automated training sessions for three new
undergraduate RAs (a senior and a junior majoring in economics, and a junior
mathematics major). In the automated training sessions, each of the three
RAs was asked to classify random subsamples of the labeled observations, and
to repeat the labeling practice until their judgments agreed with the
coauthor's at least 80\% of the time. Subsequently, each of the RAs
independently labeled the entire dataset in February--April 2021. Thus, each 
$\left( i,t\right) $ in NSW carries three labels.

\paragraph{Germany.}

Third, the same team of three RAs proceeded to label a 5\% random sample of
the German dataset in April--June 2021. In turn, these labels served as a
source of \textquotedblleft training sample\textquotedblright \ for yet
another team of three RAs (two juniors majoring in economics and a freshman
in statistics and data science). They labeled an additional 5\% random
sample in June 2021. In total, 10\% of the German data is triple-labeled.

\paragraph{Risk of \textquotedblleft Collusion\textquotedblright \ Is Low.}

In the computerized procedures for NSW and Germany, each RA is given \textit{%
one randomly selected observation for labeling} at a time. We believe the
risk of \textquotedblleft collusion\textquotedblright \ among RAs is low
because copying each other's answers would require (i) keeping records of
random sequences of thousands of observations with their station-quarter
identifiers, (ii) exchanging these long records, and (iii) matching each
other's answers across different random sequences. Such a conspiracy is
conceivable in principle but prohibitively time-consuming in practice.
Honestly labeling all observations just once would be much easier.

\paragraph{Summary Statistics.}

Table \ref{Table - Summary statistics} reports summary statistics. Based on
these manual-classification results, we define $cycle_{i,t}$ as a binary
variable indicating the presence of clear cycles. In WA, each observation is
labeled exactly once, based on the consensus of two RAs. We set $%
cycle_{i,t}=1$ if station-quarter $\left( i,t\right) $ is labeled as
\textquotedblleft cycling,\textquotedblright \ and $0$ otherwise. In the NSW
and German data, which contain more ambiguous patterns, we assigned three
RAs to label each observation individually, and hence each $\left(
i,t\right) $\ is triple-labeled. We set $cycle_{i,t}=1$ for observations
with triple \textquotedblleft cycling\textquotedblright \ labels (i.e.,
based on three RAs' unanimous decisions), and $0$ otherwise.\footnote{%
We assess the sensitivity of our results under an alternative criterion in
Appendix B.3.} Thus, we prepare the target for automatic detection in a
relatively conservative manner. 
\begin{table}[tbh]
\caption{Summary Statistics}
\label{Table - Summary statistics}
\begin{center}
\fontsize{9pt}{11pt}\selectfont%
\begin{tabular}{lccc}
\hline \hline
& $\left( 1\right) $ & $\left( 2\right) $ & $\left( 3\right) $ \\ 
Dataset & Western Australia & New South Wales & Germany \\ \hline
Sample period (yyyy/mm/dd) & $2001/1/3-2020/6/30$ & $2016/8/1-2020/7/31$ & $%
2014/6/8-2020/1/7$ \\ 
Number of gasoline stations & $821$ & $1,226$ & $14,780$ \\ 
Number of calendar quarters & $77$ & $15$ & $26$ \\ 
Number of station-quarters & $25,463$ & $9,693$ & $353,086$ \\ 
Of which: &  &  &  \\ 
Labeled as \textquotedblleft cycling\textquotedblright \ by 3 RAs & $0$ $\
\left( 0.0\% \right) $ & $6,878$ $\  \left( 71.0\% \right) $ & $14,116$ $\
\left( 39.6\% \right) $ \\ 
Labeled as \textquotedblleft cycling\textquotedblright \ by 2 RAs & $0$ $\
\left( 0.0\% \right) $ & $906$ $\  \left( 9.4\% \right) $ & $7,173$ $\  \left(
20.1\% \right) $ \\ 
Labeled as \textquotedblleft cycling\textquotedblright \ by 1 RA & $15,007$ $%
\  \left( 61.1\% \right) $ & $759$ $\  \left( 7.8\% \right) $ & $6,280$ $\
\left( 17.6\% \right) $ \\ 
Not labeled as \textquotedblleft cycling\textquotedblright \ by any RA & $%
9,562$ $\  \left( 38.9\% \right) $ & $1,150$ $\  \left( 11.9\% \right) $ & $%
8,116 $ $\  \left( 22.7\% \right) $ \\ 
Total manually labeled & $24,569$ $\  \left( 100.0\% \right) $ & $9,693$ $\
\left( 100.0\% \right) $ & $35,685$ $\  \left( 100.0\% \right) $ \\ 
Not manually labeled & $894$ & $0$ & $317,401$ \\ \hline \hline
\end{tabular}
\begin{minipage}{475pt}
{\fontsize{9pt}{9pt}\selectfont \smallskip  \textit{Note}: Each ``manually labeled'' station-quarter observation in the WA data is single-labeled as either ``cycling,'' ``maybe cycling,'' or ``not cycling,'' whereas the NSW and German data are triple-labeled. See main text for details.}
\end{minipage}
\end{center}
\end{table}

\subsection{Rationale for Daily Frequency and Quarterly Window}

Several considerations led us to use the daily sampling frequency and the
quarterly time window.

First, we prioritize setting a common time frame for all three datasets. Our
goal is to compare the performance of various methods in multiple different
datasets under the same protocol; a detailed case study of any single
region/country is not our main objective. The daily frequency is the finest
granularity that can be commonly used across all datasets because retail
prices in WA are fixed for 24 hours due to regulation (see section 3.1). It
is also the finest granularity used in most other studies (however, see
below for our discussion of the German data).

Second, cyclicality implies repetition, the identification of which requires
a sufficiently long time window. The existing studies on WA and NSW report
cycles with frequencies of one to several weeks, whereas those on Germany
report both weekly and intra-day cycles. The 12--13 weeks of a calendar
quarter permit repeated observations of relatively long (e.g., monthly)
cycles.

Third, shorter-than-daily (e.g., hourly) frequencies would be too
\textquotedblleft costly\textquotedblright \ for our research design, as
systematic manual verification is its essential component. Eyeballing and
labeling a 10\% subsample of the entire German dataset at the hourly
(instead of daily) frequency would require 24 times more labor: 480 hours $%
\times $ 24 = 11,520 hours. At the hourly wage of \$13.50, the total cost
would be \$155,520.

Fourth, we avoid longer-than-quarterly time windows for two reasons. One is
that macroeconomic factors (such as business cycles, financial crises, and
geopolitical upheavals in the world crude oil market) tend to feature
prominently in a time horizon longer than 90 days, which increases noise.
Another reason is that longer windows tend to complicate classification, as
cycles might appear in only one part of the graph but not others.

For these reasons, the daily frequency and the quarterly horizon are
suitable for our purposes. Note that our choice is driven by the comparative
research design, practical considerations, and budget constraints, not
conceptual limitations. All of the methods can be applied to time-series
data of any frequency and length in principle.

\paragraph{On Intra-Day Cycles in the German Data.}

We are aware of multiple studies that document intra-day price cycles in
Germany. The first investigation into the German retail fuel markets by
Bundeskartellamt (2011) studies data from four major cities (Hamburg,
Leipzig, Cologne, and Munich) in January 2007--June 2010 and highlights
three patterns. First, weekly cycles exist in both diesel and gasoline
prices, with the highest prices on Fridays and the lowest prices on Sundays
and Mondays. Second, intra-day cycles exist as well, with many small price
reductions during the day and fewer, larger increases in the evening. Third,
stations operated by Aral (BP) and Shell typically lead those price
increases, in which one follows the other within three hours in 90\% of the
cases, followed by three other major chains.

Given the well-documented presence of intra-day cycles, one might wonder
whether our focus on the daily data and multi-day cycles leads to an
important omission. Our answer is \textquotedblleft yes,\textquotedblright \
but this issue is orthogonal to the main purpose of this research.

By aggregating the underlying minute-by-minute data to 24-hour averages, we
lose these interesting short-run movements. Our choice of the daily
frequency is driven by the comparative design of our research, which
prioritizes the systematic comparisons across the three datasets and
(costly) manual verification. Thus, researchers who wish to conduct an
in-depth case study of the German fuel markets might want to analyze
intra-day patterns as well.

Nevertheless, the presence of shorter cycles does not preclude that of
longer cycles; Bundeskartellamt (2011) confirms the existence of both (see
above). One should also note that the intra-day cycles seem to follow a
specific time schedule in which prices (i) rapidly increase at night between
20:00 and 24:00 hours and (ii) gradually decrease from around 6:00 in the
following morning (Siekmann 2017). As Linder (2018) correctly points out,
such a deterministic pattern is more consistent with dynamic price
discrimination than Maskin and Tirole's Edgeworth cycles (recall Property
3---stochasticity---in section 2.1). Hence, while interesting, the intra-day
cycles in Germany are outside the scope of this paper.

\section{Models and Methods for Automatic Detection}

This section explains (i) how we formalize the four existing methods, (ii)
the six new methods that we propose, and (iii) the way we optimize the
parameter values of each model.

\subsection{Existing Methods Mostly Focus on Asymmetry}

The existing methods in the literature almost exclusively focus on
asymmetry. We formalize four of them as simple parametric models.

\paragraph{Method 1: Positive Runs vs. Negative Runs (\textquotedblleft
PRNR\textquotedblright ).}

Castanias and Johnson (1993) compare the lengths of positive and negative
changes. We formalize this idea by classifying each station-quarter as
cycling ($cycle_{i,t}=1$) if and only if%
\begin{equation}
mean\left( len\left( run^{+}\right) \right) <mean\left( len\left(
run^{-}\right) \right) +\theta ^{PRNR},
\label{eq - condition (positive runs vs. negative runs)}
\end{equation}%
where $len\left( run^{+}\right) $\ and $len\left( run^{-}\right) $\ denote
the lengths of consecutive (multi-day) price increases/zero changes and
decreases within quarter $t$, respectively. The means are taken over these
\textquotedblleft runs.\textquotedblright \ $\theta ^{PRNR}\approx 0$ is a
scalar threshold, which we treat as a parameter.\footnote{%
Eckert (2002) proposes a more comprehensive version of this idea, which
compares the \textit{distributions} of positive and negative runs across
lengths, by using the Kolmogorov-Smirnov test.}

\paragraph{Method 2: Mean Increase vs. Mean Decrease (\textquotedblleft
MIMD\textquotedblright ).}

Eckert (2002) compares the magnitude of the mean increase and the mean
decrease. Formally, station-quarter $\left( i,t\right) $ is cycling if and
only if%
\begin{equation}
\left \vert mean_{d\in t}\left( \Delta p_{i,d}^{+}\right) \right \vert
>\left \vert mean_{d\in t}\left( \Delta p_{i,d}^{-}\right) \right \vert
+\theta ^{MIMD},  \label{eq - condition (mean increase vs. mean decrease)}
\end{equation}%
where $\Delta p_{i,d}^{+}$ and $\Delta p_{i,d}^{-}$ denote positive and
negative daily price changes at station $i$ (between days $d$ and $d-1$),
respectively, and $\theta ^{MIMD}\approx 0$ is a scalar threshold. That is,
a cycle is detected when the average price increase is greater than the
average price decrease.\footnote{%
Eckert (2003) uses this method as well. Clark and Houde (2014) propose its
variant: the ratio of the median price increase to the median price
decrease, with 2 as a threshold to define cyclical subsamples.}

\paragraph{Method 3: Negative Median Change (\textquotedblleft
NMC\textquotedblright ).}

Lewis (2009) classifies $cycle_{i,t}=1$ if and only if%
\begin{equation}
median_{d\in t}\left( \Delta p_{i,d}\right) <\theta ^{NMC},
\label{eq - condition (negative median change)}
\end{equation}%
where $\Delta p_{i,d}$ denotes a price change between days $d$ and $d-1$,
and $\theta ^{NMC}\approx 0$ is a scalar threshold.\ In other words, the
significantly negative median change is taken as evidence of price cycles.%
\footnote{%
Many subsequent studies use this method, including Wills-Johnson and Bloch
(2010), Doyle, Muehlegger, and Samphantharak (2010), Lewis and Noel (2011),
Lewis (2012), Eckert and Eckert (2013), Zimmerman, Yun, and Taylor (2013),
and Byrne (2019). As a threshold for discretization, Lewis (2012) uses $-0.2$
US cents per gallon, whereas Doyle et al. (2010) and Zimmerman et al. (2013)
use $-0.5$ US cents per gallon.}

\paragraph{Method 4: Many Big Price Increases (\textquotedblleft
MBPI\textquotedblright ).}

Byrne and de Roos (2019) identify price cycles with the condition%
\begin{equation}
\sum_{d\in t}\mathbb{I}\left \{ \Delta p_{i,d}>\theta _{1}^{MBPI}\right \}
\geq \theta _{2}^{MBPI},  \label{eq - condition (many big price increases)}
\end{equation}%
where $\mathbb{I}\left \{ \cdot \right \} $ is an indicator function that
equals 1 if the condition inside the bracket is satisfied, and 0 otherwise. $%
\theta _{1}^{MBPI}$ and $\theta _{2}^{MBPI}$ are thresholds for
\textquotedblleft big\textquotedblright \ and \textquotedblleft
many\textquotedblright \ price increases, respectively. They set $\theta
_{1}^{MBPI}=6$ (Australian cents/liter) and $\theta _{2}^{MBPI}=3.75$ (per
quarter) in studying the WA data.\footnote{%
Lewis (2009) also uses a similar method, with $\theta _{1}^{MBPI}=4$ (US
cents/gallon) in a single day or two consecutive days.} Thus, many instances
of big price increases are taken as evidence of price cycles.

\paragraph{Other Existing Methods.}

These methods are among the most cited in the literature, but our listing is
not exhaustive. Other influential papers use a variety of methods. Let us
briefly discuss three of them. First, Noel (2007) proposes a Markov
switching model with three unobserved states, two of which correspond to
positive and negative runs, respectively, and the third corresponds to a
non-cyclical regime.\footnote{%
Because these states are modeled as unobserved objects, using this approach
as a definition is not straightforward. Zimmerman et al. (2013) propose
another definition that shares the spirit of Markov switching regressions:
(i) Compare the probability that a price increase (decrease) is observed
after two consecutive price increases (decreases); and (ii) if the
conditional probability of a third consecutive increase is smaller than that
of a third decrease, take it as an indicator of cycles. We regard their
approach as a variant of Castanias and Johnson's method. Finally, Noel
(2018) defines the relenting and undercutting phases by consecutive days
with cumulative increases and decreases of at least 3 Australian cents per
liter, respectively, which is also close to Castanias and Johnson's (1993)
idea.} Second, Deltas (2008) and many others regress retail price on
wholesale price to describe asymmetric responses. Third, Foros and Steen
(2013) regress price on days-of-week dummies to describe weekly cycles.
These papers offer valuable insights, and their methods are suitable in
their respective contexts. However, they are not specifically designed for
defining or detecting cycles.

\subsection{Our Proposals to Capture Cyclicality}

We propose six new methods. Methods 5--6 are based on spectral analysis, and
hence are attractive as formal mathematical definitions of regular cycles.
By contrast, Methods 7--8 build on nonparametric regressions and
machine-learning techniques, respectively, and are more suitable for
capturing nuanced patterns and replicating human recognition of cycles.
Methods 9--10 combine some or all of the previous methods. Appendix A.1
provides additional technical details.

\paragraph{Method 5: Fourier Transform (\textquotedblleft
FT\textquotedblright ).}

Fourier analysis is a mathematical method for detecting and characterizing
periodicity in time-series data. When a continuous function of time $g\left(
x\right) $ is sampled at regular time intervals with spacing $\Delta x$, the
sample analog of the Fourier power spectrum (or \textquotedblleft
periodogram\textquotedblright ) is%
\begin{equation}
P\left( f\right) \equiv \frac{1}{N}\left \vert \sum_{n=1}^{N}g_{n}e^{-2\pi
ifx_{n}}\right \vert ^{2},  \label{eq - classical periodogram}
\end{equation}%
where $f$ is frequency, $N$ is the sample size, $g_{n}\equiv g\left( n\Delta
x\right) $, $i\equiv \sqrt{-1}$ is the imaginary unit (not to be confused
with our gas-station index), and $x_{n}$ is the time stamp of the $n$-th
observation. It is a positive, real-valued function that quantifies the
contribution of each frequency $f$ to the time-series data $\left(
g_{n}\right) _{n=1}^{N}$.\footnote{%
Appendix A.1 (Method 5) introduces FT to readers who are not familiar with
Fourier analysis.}

We focus on the highest point of $P\left( f\right) $ and detect cycles if
and only if%
\begin{equation}
\max_{f}P_{i,t}\left( f\right) >\theta _{\max }^{FT},
\label{eq - condition A (Fourier, max)}
\end{equation}%
where $P_{i,t}\left( f\right) $ is the periodogram (\ref{eq - classical
periodogram}) of station-quarter $\left( i,t\right) $, and $\theta _{\max
}^{FT}>0$ is a scalar threshold parameter.

\paragraph{Method 6: Lomb-Scargle (\textquotedblleft LS\textquotedblright )
Periodogram.}

The LS periodogram (Lomb 1976, Scargle 1982) characterizes periodicity in
unevenly sampled time series.\footnote{%
Our data are evenly sampled at the daily frequency and can be analyzed by FT
alone, but the LS periodogram offers additional benefits. One is conceptual:
it is interpretable as a kind of nonparametric regression---see Appendix A.1
(Method 6). Another is practical: its off-the-shelf computational
implementation can offer more granular periodograms.} It has been
extensively used in astrophysics because astronomical observations are
subject to weather conditions and diurnal, lunar, or seasonal cycles.
Formally, it is a generalized version of the classical periodogram (\ref{eq
- classical periodogram}):\footnote{%
Appendix A.1 (Method 6) explains how this expression relates to FT.}%
\begin{equation}
P^{LS}\left( f\right) =\frac{1}{2}\left \{ \frac{\left( \sum_{n}g_{n}\cos
\left( 2\pi f\left[ x_{n}-\tau \right] \right) \right) ^{2}}{\sum_{n}\cos
^{2}\left( 2\pi f\left[ x_{n}-\tau \right] \right) }+\frac{\left(
\sum_{n}g_{n}\sin \left( 2\pi f\left[ x_{n}-\tau \right] \right) \right) ^{2}%
}{\sum_{n}\sin ^{2}\left( 2\pi f\left[ x_{n}-\tau \right] \right) }\right \}
,  \label{eq - Lomb-Scargle periodogram}
\end{equation}%
where $\tau $ is specified for each frequency $f$ as%
\begin{equation}
\tau =\frac{1}{4\pi f}\tan ^{-1}\left( \frac{\sum_{n}\sin \left( 4\pi
fx_{n}\right) }{\sum_{n}\cos \left( 4\pi fx_{n}\right) }\right) .
\label{eq - tau in LS periodogram}
\end{equation}%
We propose the following threshold condition to detect cycles:%
\begin{equation}
\max_{f}P_{i,t}^{LS}\left( f\right) >\theta _{\max }^{LS},
\label{eq - condition A (Lomb-Scargle, max)}
\end{equation}%
where $\theta _{\max }^{LS}>0$\ is a scalar threshold parameter.

\paragraph{Method 7: Cubic Splines (\textquotedblleft CS\textquotedblright ).%
}

This method captures cycles' frequency in a less structured manner than FT
and LS by using cubic splines. A spline is a piecewise polynomial function.
We smooth the discrete (daily) time series by interpolating it with a cubic
Hermite interpolater, which is a spline where each piece is a third-degree
polynomial of Hermite form.\footnote{%
Appendix A.1 (Method 7) explains the details of this functional form.} For
each $\left( i,t\right) $, we fit CS to its demeaned price series, $%
\overline{p}_{i,d}\equiv p_{i,d}-mean_{d\in t}\left( p_{i,d}\right) $, and
count the number of times the fitted function $\overline{CS}_{i,t}\left(
d\right) $ crosses the $d$-axis (i.e., equals 0). That is, we count the
number of real roots and detect cycles with the condition,%
\begin{equation}
\#roots\left( \overline{CS}_{i,t}\left( d\right) \right) >\theta
_{root}^{CS},  \label{eq - condition (CS roots)}
\end{equation}%
where $\theta _{root}^{CS}>0$\ is a scalar parameter. Thus, any frequent
oscillations (not limited to the sinusoidal ones as in FT or LS) become a
sign of cycles.

\paragraph{Method 8: Long Short-Term Memory (\textquotedblleft
LSTM\textquotedblright ).}

Recurrent neural networks with LSTM (Hochreiter and Schmidhuber 1997) are a
class of artificial neural network (ANN) models for sequential data. LSTM
networks have become a \textquotedblleft de-facto standard\textquotedblright
\ for recognizing and predicting complicated patterns in many applications,
including speech, handwriting, language, and polyphonic music. Because LSTM
is relatively new, we explain this method in greater detail.

Econometrically speaking, LSTM is a nonparametric model for time-series
analysis. It is a recursive dynamic model whose behavior centers on a
collection of pairs of $B_{l}\times 1$ vector-valued latent state variables, 
$\mathbf{s}_{d}^{l}$ and $\mathbf{c}_{d}^{l}$, where $l=1,2,\cdots ,L$ is an
index of layers. As this notation suggests, we use a multi-layer
architecture (a.k.a. \textquotedblleft deep\textquotedblright \ neural
networks) to enhance the model's flexibility.\footnote{%
Except for the multi-layer design, our specification mostly follows Greff,
Srivastava, Koutn\'{\i}k, Steunebrink, and Schmidhuber (2017), in which one
of the original proponents of LSTM and his team compare many of its variants
and show that their simple \textquotedblleft vanilla\textquotedblright \
specification outperforms others.} $B_{l}$ represents the number of blocks\
per layer, which are analogous to \textquotedblleft
neurons\textquotedblright \ (basic computing units) in other ANN models. $%
\mathbf{s}_{d}^{l}$\ is an output state\ that represents the current,
\textquotedblleft short-term\textquotedblright \ state, whereas $\mathbf{c}%
_{d}^{l}$\ is called a cell state\ and retains \textquotedblleft long-term
memory.\textquotedblright \ The latter is designed to capture lagged
dependence between the state and input variables, thereby playing the role
of a memory cell\ in electronic computers.

These state variables evolve according to the following Markov process:%
\begin{eqnarray}
\mathbf{s}_{d}^{l} &=&\underbrace{\tanh \left( \mathbf{c}_{d}^{l}\right) }_{%
\text{\textquotedblleft output\textquotedblright }}\  \circ \underbrace{%
\Lambda \left( \mathbf{\omega }_{1}^{l}+\mathbf{\omega }_{2}^{l}\Delta p_{d}+%
\mathbf{\omega }_{3}^{l}\mathbf{s}_{d}^{l-1}\right) }_{\text{%
\textquotedblleft output gate\textquotedblright }}\ ,\text{ and}
\label{eq - multi-layer LSTM state 1} \\
\mathbf{c}_{d}^{l} &=&\underbrace{\tanh \left( \mathbf{\omega }_{4}^{l}+%
\mathbf{\omega }_{5}^{l}\Delta p_{d}+\mathbf{\omega }_{6}^{l}\mathbf{s}%
_{d}^{l-1}\right) }_{\text{\textquotedblleft input\textquotedblright }}\
\circ \underbrace{\Lambda \left( \mathbf{\omega }_{7}^{l}+\mathbf{\omega }%
_{8}^{l}\Delta p_{d}+\mathbf{\omega }_{9}^{l}\mathbf{s}_{d}^{l-1}\right) }_{%
\text{\textquotedblleft input gate\textquotedblright }}  \notag \\
&&+\mathbf{c}_{d}^{l-1}\circ \underbrace{\left[ 1-\Lambda \left( \mathbf{%
\omega }_{7}^{l}+\mathbf{\omega }_{8}^{l}\Delta p_{d}+\mathbf{\omega }%
_{9}^{l}\mathbf{s}_{d}^{l-1}\right) \right] }_{\text{\textquotedblleft
forget gate\textquotedblright }}\ ,  \label{eq - multi-layer LSTM state 2}
\end{eqnarray}%
where $d=1,2,\cdots ,D$ is our index of days, $\Delta p_{d}\equiv
p_{d}-p_{d-1}$ (we set $\Delta p_{1}=0$), $\tanh \left( x\right) \equiv 
\frac{e^{x}-e^{-x}}{e^{x}+e^{-x}}$ is the hyperbolic tangent function, $%
\circ \ $denotes the Hadamard (element-wise) product, and$\  \Lambda \left(
x\right) \equiv \frac{e^{x}}{1+e^{x}}$ is the cumulative distribution
function (CDF) of the logistic distribution.\footnote{%
See Appendix A.1 (Method 8) for further details on this specification and
computational implementation.} The $\mathbf{\omega }$s are weight parameters
with the following dimensionality: (i) $\mathbf{\omega }_{1}^{l}$, $\mathbf{%
\omega }_{2}^{l}$, $\mathbf{\omega }_{4}^{l}$, $\mathbf{\omega }_{5}^{l}$%
\textbf{, }$\mathbf{\omega }_{7}^{l}$, and $\mathbf{\omega }_{8}^{l}$ are $%
B_{l}\times 1$ vectors; and (ii) $\mathbf{\omega }_{3}^{l}$, $\mathbf{\omega 
}_{6}^{l}$, and $\mathbf{\omega }_{9}^{l}$ are $B_{l}\times B_{l-1}$
matrices. Thus, $\mathbf{B\equiv }\left( B_{1},B_{2},\cdots ,B_{L}\right) $
determines the effective number of latent state variables and parameters,
and hence the flexibility of the model.

The first layer $l=1$ of time $d$ takes as input the states of the last
layer $l=L$ of time $d-1$. Thus, $\left( \mathbf{s}_{d}^{l-1}\mathbf{,c}%
_{d}^{l-1},B_{l-1}\right) $ in the above should be replaced by $\left( 
\mathbf{s}_{d-1}^{L}\mathbf{,c}_{d-1}^{L},B_{L}\right) $\ when $l=1$. After
the final layer $L$ of the last day $D=90$ of quarter $t$, we detect cycles
in station-quarter $\left( i,t\right) $ if and only if%
\begin{equation}
s^{\ast }\left( \mathbf{p}_{i,t};\mathbf{\theta }^{LSTM}\right) \equiv
\omega _{10}+\mathbf{\omega }_{11}^{\prime }\mathbf{s}_{D}^{L}>0,
\label{eq - condition (LSTM)}
\end{equation}%
where $\omega _{10}$ is a scalar, $\mathbf{\omega }_{11}$ is a $B_{L}\times
1 $\ vector, and $\mathbf{\theta }^{LSTM}\equiv \left( \mathbf{\omega },L,%
\mathbf{B}\right) $ collectively denotes all parameters, including (i) the
many weights in $\mathbf{\omega }\equiv \left( \left( \mathbf{\omega }%
_{1}^{l},\mathbf{\omega }_{2}^{l},\cdots ,\mathbf{\omega }_{9}^{l}\right)
_{l=1}^{L},\omega _{10},\mathbf{\omega }_{11}\right) $, (ii) the number of
layers $L$, and (iii) the profile of the number of blocks in each layer, $%
\mathbf{B}$. We set $L=3$ and $\mathbf{B}=\left( 16,8,4\right) $, and find
the value of $\mathbf{\omega }$\ that approximately maximizes the accuracy
of prediction (to be explained in section 4.3 and Appendix A.2).

In summary, LSTM sequentially processes the daily price data in a flexible
Markov model with many latent states, and uses the terminal state $s^{\ast }$%
\ as a latent score to detect cycles.

\paragraph{Method 9: Ensemble in Random Forests (\textquotedblleft
E-RF\textquotedblright ).}

This method combines Methods 1--7 within random forests (RF), which is a
class of nonparametric regressions. Let 
\begin{equation}
g_{i,t}^{m}\equiv LHS_{i,t}^{m}-RHS_{i,t}^{m}  \label{eq - gap function}
\end{equation}%
denote a \textquotedblleft gap,\textquotedblright \ the scalar difference
between the left-hand side (LHS) and the right-hand side (RHS) of the
inequality that defines each method $m=1,2,\cdots ,M$, excluding the
threshold parameter, $\mathbf{\theta }^{m}$. For example, inequality (\ref%
{eq - condition (mean increase vs. mean decrease)}) defines Method 2. Hence, 
$g_{i,t}^{2}=\left \vert mean_{d\in t}\left( \Delta p_{i,d}^{+}\right)
\right \vert -\left \vert mean_{d\in t}\left( \Delta p_{i,d}^{-}\right)
\right \vert $.\footnote{%
All of Methods 1--7 except 4 are one-parameter models like this example. For
Method 4, we define $g_{i,t}^{4}\equiv \sum_{d\in t}\mathbb{I}\left \{
\Delta p_{i,d}>\theta _{1}^{MBPI\ast }\right \} $, where $\theta
_{1}^{MBPI\ast }$ is the accuracy-maximizing value of $\theta _{1}^{MBPI}$.}
Let 
\begin{equation}
\mathbf{g}_{i,t}\equiv \left( g_{i,t}^{m}\right) _{m=1}^{M}
\label{eq - all gap functions}
\end{equation}%
denote their vector, where $M=7$.\footnote{%
Our computational implementation also incorporates two additional variants
of each of Methods 5--7, which we explain in Appendix A.1 (Method 9). Hence,
the eventual value of $M$ is $7+\left( 2\times 3\right) =13$.} We construct
a decision-tree classification algorithm that takes $\mathbf{g}_{i,t}$\ as
inputs and predicts $cycle_{i,t}=1$ if and only if%
\begin{equation}
h\left( \mathbf{g}_{i,t};\mathbf{\omega }^{RF},\mathbf{\kappa }^{RF}\right)
\equiv \sum_{k=1}^{K}\omega _{k}^{RF}\mathbb{I}\left \{ \mathbf{g}_{i,t}\in
R_{k}\right \} \equiv \sum_{k=1}^{K}\omega _{k}^{RF}\phi \left( \mathbf{g}%
_{i,t};\mathbf{\kappa }_{k}^{RF}\right) >0,
\label{eq - condition (ensemble RF)}
\end{equation}%
where $K$ is the number of adaptive basis functions, $\omega _{k}^{RF}$ is
the weight of the $k$-th basis function, $R_{k}$ is the $k$-th region in the 
$M$-dimensional space of $\mathbf{g}_{i,t}$, and $\mathbf{\kappa }_{k}^{RF}$
encodes both the choice of variables (elements of $\mathbf{g}_{i,t}$) and
their threshold values that determine region $R_{k}$.\footnote{%
See Murphy (2012, ch. 16) for an introduction to adaptive basis-function
models including RF.} Because finding the truly optimal partitioning is a
computationally difficult (combinatorial) problem, we use an RF algorithm to
stochastically approximate it.\footnote{%
See Appendix A.1 (Method 9) for further details.} Thus, this method
aggregates and generalizes Methods 1--7 in a flexible manner that permits
(i) multiple thresholds and (ii) interactions between $g_{i,t}^{m}$s. We
denote its full set of parameters by $\mathbf{\theta }^{RF}\equiv \left( 
\mathbf{\omega }^{RF},\mathbf{\kappa }^{RF}\right) \equiv \left( \left(
\omega _{k}^{RF}\right) _{k=1}^{K},\left( \mathbf{\kappa }_{k}^{RF}\right)
_{k=1}^{K}\right) $.

\paragraph{Method 10: Ensemble in LSTM (\textquotedblleft
E-LSTM\textquotedblright ).}

This method combines Methods 1--8 within an extended LSTM by incorporating $%
\mathbf{g}_{i,t}$\ in (\ref{eq - all gap functions}) as additional variables
in the laws of motion:%
\begin{eqnarray}
\mathbf{s}_{d}^{l} &=&\tanh \left( \mathbf{c}_{d}^{l}\right) \circ \Lambda
\left( \mathbf{\omega }_{1}^{l}+\mathbf{\omega }_{2}^{l}\Delta p_{d}+\mathbf{%
\omega }_{3}^{l}\mathbf{s}_{d}^{l-1}+\mathbf{\omega }_{12}^{l}\mathbf{g}%
\right) ,\text{ and}  \label{eq - multi-layer LSTM state 1 ensemble} \\
\mathbf{c}_{d}^{l} &=&\tanh \left( \mathbf{\omega }_{4}^{l}+\mathbf{\omega }%
_{5}^{l}\Delta p_{d}+\mathbf{\omega }_{6}^{l}\mathbf{s}_{d}^{l-1}+\mathbf{%
\omega }_{13}^{l}\mathbf{g}\right) \circ \Lambda \left( \mathbf{\omega }%
_{7}^{l}+\mathbf{\omega }_{8}^{l}\Delta p_{d}+\mathbf{\omega }_{9}^{l}%
\mathbf{s}_{d}^{l-1}+\mathbf{\omega }_{14}^{l}\mathbf{g}\right)  \notag \\
&&+\mathbf{c}_{d}^{l-1}\circ \left[ 1-\Lambda \left( \mathbf{\omega }%
_{7}^{l}+\mathbf{\omega }_{8}^{l}\Delta p_{d}+\mathbf{\omega }_{9}^{l}%
\mathbf{s}_{d}^{l-1}+\mathbf{\omega }_{14}^{l}\mathbf{g}\right) \right] ,
\label{eq - multi-layer LSTM state 2 ensemble}
\end{eqnarray}%
where $\left( \mathbf{\omega }_{12}^{l},\mathbf{\omega }_{13}^{l},\mathbf{%
\omega }_{14}^{l}\right) $ are $B_{l}\times M$ matrices of weight parameters
for $\mathbf{g}_{i,t}$ (we suppress $\left( i,t\right) $ subscript here).
Other implementation details are the same as Method 8.

\subsection{Optimization of Parameter Values (\textquotedblleft
Training\textquotedblright )}

\paragraph{Accuracy Maximization.}

Whereas the existing research typically calibrates (i.e., manually tunes)
the threshold parameters, we optimize this process by choosing the parameter
values that maximize accuracy, which we define as the percentage of correct
predictions,%
\begin{equation}
\% \text{ correct}\left( \mathbf{\theta }\right) \equiv \frac{\sum_{\left(
i,t\right) }\mathbb{I}\left \{ \widehat{cycle}_{i,t}\left( \mathbf{\theta }%
\right) =cycle_{i,t}\right \} }{\# \text{ }all\text{ }predictions}\times 100,
\label{eq - percent correct}
\end{equation}%
where $\widehat{cycle}_{i,t}\left( \mathbf{\theta }\right) \in \left \{
0,1\right \} $ is the algorithmic prediction for observation $\left(
i,t\right) $ at parameter value $\mathbf{\theta }$, and $cycle_{i,t}\in
\left \{ 0,1\right \} $ is the manual classification label (data). We
analogously define two types of prediction errors, \textquotedblleft false
negative\textquotedblright \ and \textquotedblleft false
positive,\textquotedblright \ in Appendix A.2. Thus,%
\begin{equation}
\mathbf{\theta }^{\ast }\equiv \arg \max_{\mathbf{\theta }}\text{ \ }\% 
\text{ correct}\left( \mathbf{\theta }\right)
\label{eq - accuracy maximizing theta}
\end{equation}%
characterizes the optimized\ (or \textquotedblleft trained\textquotedblright
)\ model for each method.\footnote{%
See Appendix A.2 for further details.}

\paragraph{Splitting Data into Training and Testing Subsamples.}

We optimize and evaluate each method as follows, separately for each of the
three datasets (WA, NSW, and Germany):

\begin{enumerate}
\item Randomly split each labeled dataset into an 80\% \textquotedblleft
training\textquotedblright \ subsample and a 20\% \textquotedblleft
testing\textquotedblright \ subsample.

\item Optimize the parameter values of each model in the 80\% training
subsample.

\item Assess its \textquotedblleft out-of-sample\textquotedblright \
prediction accuracy in the 20\% testing subsample.\footnote{%
This cross-validation procedure is particularly important for the
nonparametric models of Methods 8--10, which contain many parameters and
could potentially \textquotedblleft over-fit\textquotedblright \ the
training subsample.}

\item Repeat these three steps 101 times.\footnote{%
An odd number of bootstrap sample-splits facilitates the selection of the
medians in step 5.}

\item Report the medians of the optimized parameter values, as well as the
medians and standard deviations of the prediction-accuracy results.
\end{enumerate}

\section{Results}

Table \ref{Table - Accuracy (main)} summarizes the performance of all
methods for each dataset. We report the median accuracy, the composition of
correct and incorrect predictions, and the associated parameter value(s), $%
\mathbf{\theta }^{\ast }$, for each method.

\begin{table}[tbh]
\caption{Performance of Automatic Detection Methods}
\label{Table - Accuracy (main)}
\begin{center}
\fontsize{9pt}{11pt}\selectfont%
\begin{tabular}{lcccccccccc}
\hline \hline
& $(1)$ & $(2)$ & $(3)$ & $(4)$ & $(5)$ & $(6)$ & $(7)$ & $\left( 8\right) $
& $\left( 9\right) $ & $\left( 10\right) $ \\ 
Method & PRNR & MIMD & NMC & MBPI & FT & LS & CS & LSTM & E-RF & E-LSTM \\ 
\hline
\multicolumn{1}{c}{} &  &  &  &  &  &  &  &  &  &  \\ 
\multicolumn{11}{c}{\textit{I. Western Australia} (\# manually labeled
observations: $24,569$)} \\ 
\multicolumn{1}{c}{} &  &  &  &  &  &  &  &  &  &  \\ 
Parameter 1 & $-1.16$ & $6.13$ & $-0.20$ & $5.05$ & $0.12$ & $0.21$ & $22.50$
& $-$ & $-$ & $-$ \\ 
Parameter 2 & $-$ & $-$ & $-$ & $5$ & $-$ & $-$ & $-$ & $-$ & $-$ & $-$ \\ 
Accuracy rank & $5$ & $4$ & $9$ & $6$ & $8$ & $7$ & $10$ & $1$ & $3$ & $1$
\\ 
\% correct (median) & $90.80$ & $91.27$ & $89.34$ & $90.23$ & $90.11$ & $%
90.15$ & $85.47$ & $99.25$ & $99.04$ & $99.25$ \\ 
(Standard deviations) & $\left( 0.37\right) $ & $\left( 0.38\right) $ & $%
\left( 0.38\right) $ & $\left( 0.36\right) $ & $\left( 0.40\right) $ & $%
\left( 0.36\right) $ & $\left( 0.45\right) $ & $\left( 0.18\right) $ & $%
\left( 0.15\right) $ & $\left( 0.14\right) $ \\ 
of which cycling & $55.27$ & $55.70$ & $57.08$ & $60.74$ & $58.24$ & $57.92$
& $56.41$ & $60.62$ & $60.97$ & $60.34$ \\ 
of which not & $35.53$ & $35.57$ & $32.25$ & $29.49$ & $31.87$ & $32.23$ & $%
29.06$ & $38.62$ & $38.07$ & $38.91$ \\ 
\% false negative & $5.27$ & $5.27$ & $3.34$ & $0.71$ & $2.48$ & $3.30$ & $%
5.29$ & $0.35$ & $0.61$ & $0.31$ \\ 
\% false positive & $3.93$ & $3.46$ & $7.33$ & $9.06$ & $7.41$ & $6.55$ & $%
9.24$ & $0.41$ & $0.35$ & $0.45$ \\ 
\multicolumn{1}{c}{} &  &  &  &  &  &  &  &  &  &  \\ 
\multicolumn{11}{c}{\textit{II. New South Wales} (\# manually labeled
observations: $9,693$)} \\ 
\multicolumn{1}{c}{} &  &  &  &  &  &  &  &  &  &  \\ 
Parameter 1 & $4.20$ & $5.76$ & $1.01$ & $14.90$ & $0.20$ & $0.57$ & $4.50$
& $-$ & $-$ & $-$ \\ 
Parameter 2 & $-$ & $-$ & $-$ & $2$ & $-$ & $-$ & $-$ & $-$ & $-$ & $-$ \\ 
Accuracy rank & $7$ & $8$ & $10$ & $4$ & $6$ & $5$ & $9$ & $2$ & $3$ & $1$
\\ 
\% correct (median) & $78.55$ & $78.39$ & $70.96$ & $81.59$ & $80.71$ & $%
80.82$ & $73.90$ & $89.63$ & $87.42$ & $90.30$ \\ 
(Standard deviations) & $\left( 0.85\right) $ & $\left( 0.88\right) $ & $%
\left( 0.97\right) $ & $\left( 0.86\right) $ & $\left( 0.80\right) $ & $%
\left( 0.80\right) $ & $\left( 0.89\right) $ & $\left( 0.67\right) $ & $%
\left( 0.69\right) $ & $\left( 0.67\right) $ \\ 
of which cycling & $67.04$ & $65.09$ & $70.96$ & $64.62$ & $66.53$ & $66.43$
& $70.40$ & $67.20$ & $67.10$ & $65.60$ \\ 
of which not & $11.50$ & $13.31$ & $0.00$ & $16.97$ & $14.18$ & $14.39$ & $%
3.51$ & $22.43$ & $20.32$ & $24.70$ \\ 
\% false negative & $3.30$ & $4.85$ & $0.00$ & $6.55$ & $5.47$ & $4.02$ & $%
0.77$ & $4.33$ & $8.35$ & $2.99$ \\ 
\% false positive & $18.15$ & $16.76$ & $29.04$ & $11.86$ & $13.82$ & $15.16$
& $25.32$ & $6.03$ & $4.23$ & $6.70$ \\ 
\multicolumn{1}{c}{} &  &  &  &  &  &  &  &  &  &  \\ 
\multicolumn{11}{c}{\textit{III. Germany} (\# manually labeled observations: 
$35,685$)} \\ 
\multicolumn{1}{c}{} &  &  &  &  &  &  &  &  &  &  \\ 
Parameter 1 & $-3.48$ & $0.30$ & $-0.45$ & $1.25$ & $0.24$ & $0.62$ & $24.50$
& $-$ & $-$ & $-$ \\ 
Parameter 2 & $-$ & $-$ & $-$ & $14$ & $-$ & $-$ & $-$ & $-$ & $-$ & $-$ \\ 
Accuracy rank & $9$ & $6$ & $7$ & $5$ & $8$ & $10$ & $4$ & $3$ & $2$ & $1$
\\ 
\% correct (median) & $60.38$ & $60.61$ & $60.53$ & $65.39$ & $60.50$ & $%
60.36$ & $71.28$ & $74.61$ & $76.14$ & $79.58$ \\ 
(Standard deviations) & $\left( 0.49\right) $ & $\left( 0.50\right) $ & $%
\left( 0.52\right) $ & $\left( 0.52\right) $ & $\left( 0.56\right) $ & $%
\left( 0.59\right) $ & $\left( 0.42\right) $ & $\left( 0.44\right) $ & $%
\left( 1.46\right) $ & $\left( 0.53\right) $ \\ 
of which cycling & $0.00$ & $1.25$ & $0.07$ & $14.77$ & $0.00$ & $0.00$ & $%
25.88$ & $23.46$ & $23.96$ & $29.96$ \\ 
of which not & $60.38$ & $59.37$ & $60.46$ & $50.62$ & $60.50$ & $60.36$ & $%
45.40$ & $51.16$ & $52.18$ & $49.63$ \\ 
\% false negative & $39.62$ & $38.07$ & $39.40$ & $24.65$ & $39.50$ & $39.57$
& $14.28$ & $15.99$ & $15.75$ & $9.50$ \\ 
\% false positive & $0.00$ & $1.32$ & $0.07$ & $9.96$ & $0.00$ & $0.07$ & $%
14.45$ & $9.40$ & $8.11$ & $10.91$ \\ \hline \hline
\end{tabular}
\begin{minipage}{475pt}
{\fontsize{9pt}{9pt}\selectfont \smallskip  \textit{Note}: See section 4 for the definition of each method. Appendix B.1 (Table \ref{Table - Accuracy (additional results)}) reports additional results for the variants of Methods 5--7. Columns (8)--(10) do not report parameter values because they contain too many parameters to be listed. We randomly split the sample into an 80\%  training subsample and a 20\% testing subsample 101 times. In each split, the former subsample is used for setting parameter values, the medians of which are reported here. The accuracy statistics are also the medians from the 101 testing subsamples.}
\end{minipage}
\end{center}
\end{table}

\paragraph{WA.}

Panel I shows the results in WA, where clear-cut cycles of deterministic
frequencies are known to exist. Almost all methods achieve high accuracy
near or above 90\%. The flexible, nonparametric models of Methods 8--10 do
particularly well with above 99\% accuracy.

Some of the parameter values are informative about the underlying data
patterns. For example, CS lags behind all other methods with (a still
respectable) 85\% accuracy. Its parameter value, $\theta _{roots}^{CS}=22.5$%
, suggests the model is trained to focus on shorter cycles with frequencies
less than $90\div \frac{22.5}{2}=8$ days. Byrne and de Roos (2019) show both
weekly and two-weekly cycles exist in WA. Thus, the inferior performance of
CS stems from missing the latter, longer cycles.

Another interesting result concerns MBPI, which achieves 90\% accuracy.
Byrne and de Roos (2019) set $\theta _{1}^{MBPI}=6$ and $\theta
_{2}^{MBPI}=3.75$ in their original study of WA. Our accuracy-maximizing
values ($5.05$ and $5$, respectively) turn out to be reasonably close to
their calibrated values. This comparison illustrates how experienced
researchers' parameter tuning could approximate the results of systematic
numerical optimization. One can also interpret this finding as an external
validation of our manual classification. Given the similar parameter values
and the high accuracy, it follows that our manual classification must be
broadly consistent with Byrne and de Roos's eyeballing results.

\paragraph{NSW.}

Panel II reports the results in NSW. Cycle detection in NSW is not as easy
as in WA, but most methods achieve near or above 80\% accuracy. The
nonparametric methods are top performers again (87\%--90\%), followed by
MBPI and the spectral methods (81\%--82\%). By contrast, CS (74\%) and NMC
(71\%) make mostly degenerate predictions in which they classify virtually
all observations as cycles.

The poor performance of NMC is surprising in three ways. First, it performed
well in WA. Second, it is one of the most widely used methods in the
literature. Third, other methods that similarly focus on asymmetry (PRNR and
MIMD) do significantly better (78\%--79\%). This finding alone does not
necessarily invalidate the use of NMC in other datasets but cautions against
overly relying on any single metric.

\paragraph{Germany.}

Panel III shows most methods fail in Germany, where cycles are more subtle
and data are noisier (i.e., our RAs reach unanimous decisions less often).%
\footnote{%
As Table \ref{Table - Summary statistics} shows, $71\%$ of the NSW data is
unanimously labeled as \textquotedblleft cycling\textquotedblright \ by
three RAs, whereas $9.4\%+7.8\%=17.2\%$ is labeled as such by only two or
one RAs. In the German sample, only $39.6\%$ is unanimously
\textquotedblleft cycling,\textquotedblright \ whereas RAs disagree in $%
20.1\%+17.6\%=37.7\%$ of the data. Appendix B.3 reports results based on
\textquotedblleft cleaner\textquotedblright \ subsamples that eliminate such
observations with disagreements.} E-LSTM is the only method that achieves
accuracy near 80\%, followed by E-RF (76\%) and LSTM\ (75\%). Somewhat
surprisingly, CS (71\%) outperforms all other parametric models; MBPI (65\%)
is the only existing method with non-degenerate predictions, presumably
because it does not exclusively rely on asymmetry.

This profile of success and failure is intriguing. The methods that
exclusively focus on asymmetry (Methods 1--3) and deterministic cycles
(Methods 5--6) fail, whereas those that capture cyclicality in
\textquotedblleft fuzzier\textquotedblright \ manners (Methods 4 and 7)
manage to make at least some correct (non-degenerate) predictions. These
results suggest that not all of the German cycles conform to the idealized
patterns of asymmetry or cyclicality and that less rigid classification
rules could be relatively more robust to irregular patterns and noise.

The parameter values of CS ($\theta _{roots}^{CS}=24.50$) and MBPI ($\theta
_{2}^{MBPI}=14$) suggest that the German cycles are approximately weekly.\
That is, $\theta _{roots}^{CS}=24.50$ means at least as many ups and downs
are often recorded in \textquotedblleft cycling\textquotedblright \
observations, which translate into the frequency of $90\div \frac{24.5}{2}%
=7.3$ days\ or shorter. Likewise, $\theta _{2}^{MBPI}=14$ requires at least
as many \textquotedblleft big jumps\textquotedblright \ within a calendar
quarter and hence implies the frequency of $90\div 14=6.4$ days or shorter.
These numbers provide another opportunity for external validation: the
detailed case study by Bundeskartellamt (2011) confirms the presence of
weekly cycles (see section 3.3).

\paragraph{Summary.}

In summary, four findings emerge from Table \ref{Table - Accuracy (main)}.
First, the four existing methods (Methods 1--4) work well in the clean data
environments of Australia, but mostly fail in the noisier data from Germany.
The spectral methods (Methods 5--6) show similar performance. Second, by
contrast, CS (Method 7) underperforms most other methods when cycles are
clear and regular, but does relatively well in noisier cases. Third, LSTM
(Method 8) is sufficiently flexible to capture both clear and noisy cycles:
the most accurate stand-alone method. Fourth, the ensemble methods (Methods
9--10) effectively leverage the information content of Methods 1--8 and
usually outperform all of them. The fact that E-RF performs so well is
particularly interesting because it simply aggregates the descriptive
statistics from Methods 1--7 in a more flexible manner (i.e., permitting
their interactions and multiple thresholds).\clearpage

\section{How Much Data Do We Need?}

The accuracy \textquotedblleft horse racing\textquotedblright \ in the
previous section shows that more flexible methods tend to outperform simple
parametric ones, which is not surprising. The real question is the cost of
\textquotedblleft training\textquotedblright \ complicated machine-learning
algorithms, which are known to require a lot of data. This section
investigates the cost-accuracy trade-offs of the 10 methods.

\begin{figure}[htb!!!!]\centering%
\caption{Gains from Additional Data}%
\includegraphics[width=0.55\textwidth]{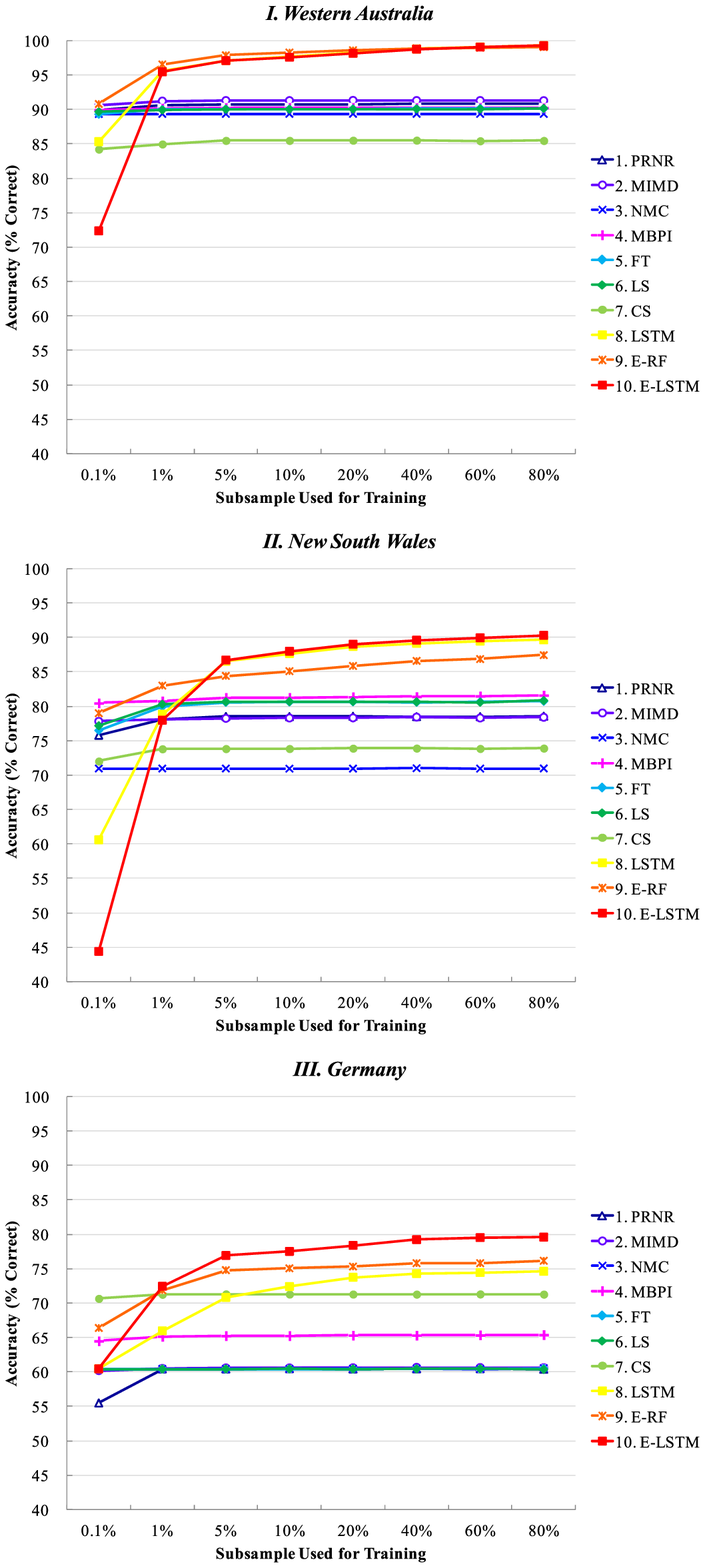}
\caption*{\footnotesize {\textit{Note}: The exact numbers
underlying these plots are reported in Panel (A) of Table \ref{Table -
Return on data} in Appendix B.2.}}%
\label{Figure - Gains from Data}
\end{figure}

The accuracy of cycle detection naturally improves with the size of the
training dataset. The rate of improvement is different across methods,
however. Figure \ref{Figure - Gains from Data} shows performance when we
restrict the training dataset to only $0.1\%$, $1\%$, $5\%$, $10\%$, $\cdots 
$, $80\%$ of the available samples.

Methods 1--7 and 9 perform surprisingly well with only $0.1\%$ of the data,
which corresponds to $25$, $10$, and $36$ observations in WA, NSW, and
Germany, respectively. The labor cost of human-generated labels is
negligible for such small samples (US$\$3.51$, US$\$2.84$, and US$\$6.48$,
respectively, based on the hourly wage of US\$13.50 for undergraduate RA
work at Yale University as of 2021). These methods are extremely cost
effective.

The fact that simple models with one or two parameters (Methods 1--7)
require only a few dozen observations is not surprising. All we have to do
is to adjust one or two numerical thresholds to distinguish cycles from
non-cycles. However, the finding that E-RF (Method 9) is equally cheap 
\textit{is} surprising. It is a highly nonlinear machine-learning model with
potentially many thresholds and interactions. This result suggests that the
building blocks of E-RF---the summary statistics derived from Methods
1--7---contain genuinely useful information that those stand-alone methods
under-utilize.

Methods 8 and 10 contain a few thousand parameters and obviously need more
data. For instance, E-LSTM's accuracy in NSW is below $50\%$ when it uses
only $10$ observations ($0.1\%$ subsamples). Fortunately, their performance
dramatically improves with a mere $1\%$ subsample, and they start
outperforming all other methods when $5\%$ subsamples are used.\footnote{%
Strictly speaking, E-RF slightly outperforms E-LSTM in subsamples up to $%
40\% $ in WA, although their mean differences are small relative to their
standard deviations (see Tables {\footnotesize {\ref{Table - Return on data}}%
}\ and {\footnotesize {\ref{Table - Return on data (stdev)}}}\ in Appendix
B.2).} The \textquotedblleft critical\textquotedblright \ sample size above
which they perform the best is in the order of several hundred observations.
The associated cost of manual labeling is only tens of RA hours, or a few
hundred US dollars.\footnote{%
Panel (B) of Table {\footnotesize {\ref{Table - Return on data}}} in
Appendix B.2 reports the total cost of manual labeling for each dateset. The
reason only \textquotedblleft several hundred observations\textquotedblright
\ are sufficient to approximately optimize \textquotedblleft a few thousand
parameters\textquotedblright \ is because various forms of regularization
restrict the effective parameter space.} Thus, even though LSTM and E-LSTM
require more data for a given accuracy level, their total cost is
surprisingly low, making them the highest-accuracy methods within a limited
amount of resources.

This finding is unexpected, but is definitely good news: state-of-the-art
machine-learning algorithms turn out to be not only useful, but also
affordable in the context of detecting Edgeworth cycles. Our conjecture is
that the cyclical patterns that humans recognize are relatively simple after
all, even though explicitly articulating them might be difficult.

\clearpage

\section{Economic and Policy Implications}

The suspicion that price cycles might be related to collusive business
practices has led many researchers and governments to collect and scrutinize
large amounts of data on fuel markets. Some papers find that the presence of
cycles is positively correlated with retail prices and markups, whereas
others find the opposite relationships (see section 1). This section
investigates how such findings depend on the definition of cycles.

\subsection{Cycles and Margins}

\paragraph{Human-Recognized Cyclicality and Margins.}

Table \ref{Table - Margins and cycles} compares the retail-wholesale margins
between \textquotedblleft cycling\textquotedblright \ and \textquotedblleft
non-cycling\textquotedblright \ observations. Column (0) is based on our
manual classification and serves as a \textquotedblleft ground
truth\textquotedblright \ benchmark. The mean margins in cycling and
non-cycling observations in WA are A\textcent 11.86 and A\textcent 9.47,
respectively. The mean difference is A\textcent 2.39. The $t$ test (based on
Welch's $t$ statistic) rejects the null hypothesis that the difference in
means is zero at the $0.1\%$ significance level. Hence, price cycles are
positively correlated with margins in WA. The same analysis yields similar
results in NSW.

However, the pattern is reversed in Germany, where margins are \textit{lower}
in cycling station-quarters. Thus, in general, the presence of cycles (as
recognized by human eyes) could be either positively or negatively
correlated with margins, depending on regions/countries.\footnote{%
Determining the exact source of heterogeneity is beyond the scope of this
paper. There can be many reasons and Edgeworth cycles are only one of the
possible mechanisms. Our purpose is to illustrate with concrete examples how
different methods could lead to different findings and policy implications.}

\begin{table}[tbh]
\caption{Profit Margins by Cycle Status}
\label{Table - Margins and cycles}
\begin{center}
\fontsize{9pt}{11pt}\selectfont%
\begin{tabular}{lccccccccccc}
\hline \hline
& $(0)$ & $(1)$ & $(2)$ & $(3)$ & $(4)$ & $(5)$ & $(6)$ & $(7)$ & $(8)$ & $%
\left( 9\right) $ & $\left( 10\right) $ \\ 
Method & Manual & PRNR & MIMD & NMC & MBPI & FT & LS & CS & LSTM & E-RF & 
E-LSTM \\ \hline
\multicolumn{1}{c}{} &  &  &  &  &  &  &  &  &  &  &  \\ 
\multicolumn{12}{c}{\textit{I. Western Australia} (\# manually labeled
observations: $24,569$)} \\ 
Cycling &  &  &  &  &  &  &  &  &  &  &  \\ 
\# obs. & $15,007$ & $14,462$ & $14,620$ & $16,147$ & $16,941$ & $16,223$ & $%
15,774$ & $15,953$ & $15,011$ & $14,994$ & $14,999$ \\ 
Mean & $11.86$ & $12.07$ & $12.21$ & $11.66$ & $11.46$ & $11.88$ & $12.03$ & 
$11.78$ & $11.86$ & $11.86$ & $11.86$ \\ 
Std. dev. & $4.01$ & $3.80$ & $3.74$ & $3.98$ & $4.13$ & $3.87$ & $3.85$ & $%
4.04$ & $4.01$ & $4.01$ & $4.01$ \\ 
Not cycling &  &  &  &  &  &  &  &  &  &  &  \\ 
\# obs. & $9,562$ & $10,107$ & $9,949$ & $8,422$ & $7,628$ & $8,346$ & $%
8,795 $ & $8,616$ & $9,558$ & $9,575$ & $9,570$ \\ 
Mean & $9.47$ & $9.30$ & $9.05$ & $9.52$ & $9.73$ & $9.08$ & $8.94$ & $9.35$
& $9.47$ & $9.47$ & $9.47$ \\ 
Std. dev. & $4.97$ & $5.04$ & $4.98$ & $5.22$ & $5.20$ & $5.18$ & $5.03$ & $%
5.02$ & $4.97$ & $4.97$ & $4.96$ \\ 
Difference &  &  &  &  &  &  &  &  &  &  &  \\ 
Mean diff. & $2.39$ & $2.77$ & $3.16$ & $2.14$ & $1.73$ & $2.80$ & $3.09$ & $%
2.43$ & $2.39$ & $2.39$ & $2.39$ \\ 
Welch's $t$ & $39.53$ & $46.74$ & $53.80$ & $32.96$ & $25.64$ & $43.53$ & $%
50.02$ & $38.67$ & $39.53$ & $39.55$ & $39.60$ \\ 
D. F. & $17,247$ & $17,771$ & $17,314$ & $13,648$ & $12,134$ & $13,263$ & $%
14,608$ & $14,723$ & $17,236$ & $17,282$ & $17,295$ \\ 
$p$ value & $<.001$ & $<.001$ & $<.001$ & $<.001$ & $<.001$ & $<.001$ & $%
<.001$ & $<.001$ & $<.001$ & $<.001$ & $<.001$ \\ 
\multicolumn{1}{c}{} &  &  &  &  &  &  &  &  &  &  &  \\ 
\multicolumn{12}{c}{\textit{II. New South Wales} (\# manually labeled
observations: $9,693$)} \\ 
Cycling &  &  &  &  &  &  &  &  &  &  &  \\ 
\# obs. & $6,878$ & $8,324$ & $8,038$ & $9,693$ & $7,303$ & $7,704$ & $7,994$
& $9,253$ & $7,052$ & $6,961$ & $7,183$ \\ 
Mean & $12.03$ & $11.73$ & $12.35$ & $11.66$ & $12.48$ & $11.76$ & $11.81$ & 
$11.58$ & $12.19$ & $12.07$ & $12.13$ \\ 
Std. dev. & $5.51$ & $5.80$ & $5.58$ & $6.04$ & $5.48$ & $5.89$ & $5.84$ & $%
5.99$ & $5.54$ & $5.53$ & $5.56$ \\ 
Not cycling &  &  &  &  &  &  &  &  &  &  &  \\ 
\# obs. & $2,815$ & $1,369$ & $1,655$ & $0$ & $2,390$ & $1,989$ & $1,699$ & $%
440$ & $2,641$ & $2,732$ & $2,510$ \\ 
Mean & $10.76$ & $11.25$ & $8.33$ & $-$ & $9.18$ & $11.28$ & $10.97$ & $%
13.48 $ & $10.25$ & $10.64$ & $10.33$ \\ 
Std. dev. & $7.10$ & $7.31$ & $7.01$ & $-$ & $6.92$ & $6.56$ & $6.85$ & $%
6.79 $ & $7.01$ & $7.08$ & $7.08$ \\ 
Difference &  &  &  &  &  &  &  &  &  &  &  \\ 
Mean diff. & $1.27$ & $0.48$ & $4.02$ & $-$ & $3.30$ & $0.48$ & $0.84$ & $%
-1.90$ & $1.94$ & $1.43$ & $1.80$ \\ 
Welch's $t$ & $8.50$ & $2.31$ & $21.94$ & $-$ & $21.24$ & $2.97$ & $4.70$ & $%
-5.76$ & $12.80$ & $9.48$ & $11.55$ \\ 
D. F. & $4,266$ & $1,663$ & $2,106$ & $-$ & $3,423$ & $2,870$ & $2,252$ & $%
472$ & $3,939$ & $4,103$ & $3,648$ \\ 
$p$ value & $<.001$ & $.021$ & $<.001$ & $-$ & $<.001$ & $.003$ & $<.001$ & $%
<.001$ & $<.001$ & $<.001$ & $<.001$ \\ 
\multicolumn{1}{c}{} &  &  &  &  &  &  &  &  &  &  &  \\ 
\multicolumn{12}{c}{\textit{III. Germany} (\# manually labeled observations: 
$35,685$)} \\ 
Cycling &  &  &  &  &  &  &  &  &  &  &  \\ 
\# obs. & $14,116$ & $0$ & $1,013$ & $72$ & $8,763$ & $7$ & $7$ & $14,281$ & 
$11,762$ & $13,574$ & $15,299$ \\ 
Mean & $98.18$ & $-$ & $99.57$ & $99.67$ & $98.73$ & $114.11$ & $115.64$ & $%
98.19$ & $98.38$ & $98.16$ & $98.18$ \\ 
Std. dev. & $3.57$ & $-$ & $6.96$ & $3.26$ & $3.84$ & $32.10$ & $31.40$ & $%
3.60$ & $3.60$ & $3.59$ & $3.51$ \\ 
Not cycling &  &  &  &  &  &  &  &  &  &  &  \\ 
\# obs. & $21,569$ & $35,685$ & $34,672$ & $35,613$ & $26,922$ & $35,678$ & $%
35,678$ & $21,404$ & $23,923$ & $22,111$ & $20,386$ \\ 
Mean & $98.65$ & $98.46$ & $98.43$ & $98.46$ & $98.38$ & $98.46$ & $98.46$ & 
$98.65$ & $98.50$ & $98.65$ & $98.68$ \\ 
Std. dev. & $4.37$ & $4.08$ & $3.96$ & $4.08$ & $4.15$ & $4.05$ & $4.05$ & $%
4.36$ & $4.30$ & $4.34$ & $4.45$ \\ 
Difference &  &  &  &  &  &  &  &  &  &  &  \\ 
Mean diff. & $-0.47$ & $-$ & $1.14$ & $1.21$ & $0.35$ & $15.65$ & $17.18$ & $%
-0.46$ & $-0.12$ & $-0.49$ & $-0.50$ \\ 
Welch's $t$ & $-11.11$ & $-$ & $5.19$ & $3.14$ & $7.26$ & $1.29$ & $1.45$ & $%
-10.86$ & $-2.77$ & $-11.55$ & $-11.86$ \\ 
D. F. & $33,984$ & $-$ & $1,031$ & $71$ & $15,941$ & $6$ & $6$ & $34,110$ & $%
27,415$ & $32,697$ & $35,595$ \\ 
$p$ value & $<.001$ & $-$ & $<.001$ & $.002$ & $<.001$ & $.245$ & $.197$ & $%
<.001$ & $.006$ & $<.001$ & $<.001$ \\ \hline \hline
\end{tabular}
\begin{minipage}{475pt}
{\fontsize{9pt}{9pt}\selectfont \smallskip  \textit{Note}: Columns (1)--(10) use the median-accuracy version of each method in Table \ref{Table - Accuracy (main)}. The unit of measurement (of means and standard deviations) is the Australian cent in WA and NSW, and the euro cent in Germany, respectively. The $p$ value indicates the probability that the difference in means is zero based on Welch's $t$ statistic and the approximate degrees of freedom.}
\end{minipage}
\end{center}
\end{table}

\paragraph{Algorithmic Cycle Detection and Margins.}

Columns (1)--(10) report the same analysis based on the 10 algorithmic
methods. In WA, all methods reach the same conclusion that margins are
higher in cycling observations. Broadly similar results also emerge in NSW,
even though one method fails (Method 3) and one reaches the opposite
conclusion (Method 7). These discrepancies suggest that researchers find a
positive or negative cycle-margin relationship depending on the operational
definition of cycles.

Our analysis of the German data highlights this point even more vividly.
Both the manual classification and Methods 7--10 suggest significantly
negative relationships between cycles and margins, but Methods 2--6 lead to 
\textit{positive} mean differences. These positive differences are highly
statistically significant in Methods 2--4. Some of them entail degenerate
predictions (see section 5), but Method 4 features reasonable parameter
values and achieves at least $65\%$ accuracy. Hence, we cannot dismiss these
discrepancies as purely random anomalies.

\paragraph{Margins and \textit{Asymmetric} Cycles.}

Note that our classification so far has focused on cyclicality but not
asymmetry. One might wonder whether our findings could change if we study 
\textit{asymmetric} cycles specifically. The answer is \textquotedblleft
no.\textquotedblright \ The results are virtually the same when we focus on
asymmetric cycles.

\begin{table}[tbh]
\caption{Profit Margins by Cycle Status and Asymmetry}
\label{Table - Margins and cycles with asymmetry}
\begin{center}
\fontsize{9pt}{11pt}\selectfont%
\begin{tabular}{lcc}
\hline \hline
& $(0)$ & $(00)$ \\ 
Method & Manual & Manual + asymmetry \\ \hline
\multicolumn{1}{c}{} &  &  \\ 
\multicolumn{3}{c}{\textit{III. Germany}} \\ 
Cycling &  &  \\ 
\# obs. & $14,116$ & $4,265$ \\ 
Mean & $98.18$ & $98.01$ \\ 
Std. dev. & $3.57$ & $3.39$ \\ 
Not cycling &  &  \\ 
\# obs. & $21,569$ & $31,420$ \\ 
Mean & $98.65$ & $98.53$ \\ 
Std. dev. & $4.37$ & $4.16$ \\ 
Difference &  &  \\ 
Mean diff. & $-0.47$ & $-0.52$ \\ 
Welch's $t$ & $-11.11$ & $-9.05$ \\ 
D. F. & $33,984$ & $6,153$ \\ 
$p$ value & $<.001$ & $<.001$ \\ \hline \hline
\end{tabular}
\begin{minipage}{475pt}
{\fontsize{9pt}{9pt}\selectfont \smallskip  \textit{Note}: Column (0) is the same as in Table \ref{Table - Margins and cycles}, which Column (00) refines by asymmetry based on negative median change. The unit of measurement (of means and standard deviations) is the euro cent. The $p$ value indicates the probability that the difference in means is zero based on Welch's $t$ statistic and the approximate degrees of freedom.}
\end{minipage}
\end{center}
\end{table}

Table \ref{Table - Margins and cycles with asymmetry} compares the mean
differences of margins based on our manual benchmark (copied from column 0
of Table \ref{Table - Margins and cycles}) and its refined version in which
we further require \textquotedblleft asymmetry\textquotedblright \ based on
the negative median change, $median_{d\in t}\left( \Delta p_{i,d}\right) <0$%
, as an additional criterion for (Edgeworth) cycles. The results are similar
both qualitatively and quantitatively.

\bigskip

In summary, the choice of the detection method could lead to qualitatively
different results and dictate the policy implications of empirical research
on Edgeworth cycles.

\subsection{Additional Findings}

The results in sections 5, 6, and 7.1 constitute our main findings, but the
curious patterns in section 7.1 present additional puzzles. We address them
in the following and report supporting evidence in Appendix C.

\paragraph{1. Why Existing Methods Work in Australia But Fail in Germany.}

Most of the cycles in Australia follow specific (almost deterministic)
frequencies and exhibit strong asymmetry, whereas German cycles are noisier
and not always asymmetric (see supplementary plots in Appendix C.1). The
existence of asymmetric \textit{non}-cycles in Germany further complicates
the issue. Hence, asymmetry-based methods correctly identify cycles in
Australia but not in Germany.

\paragraph{2. Why Margins And Cycles Correlate Positively in Australia But
Negatively in Germany.}

In all datasets, the mean and the standard deviation of margins are
positively correlated. That is, higher markups tend to accompany higher
volatility. The reason is that retail and wholesale prices are relatively
close so that the only direction in which margins can move significantly is 
\textit{upward} (unless stations are willing to incur losses). We find
volatility and cyclicality are correlated positively in Australia but
negatively in Germany (see Appendix C.2 for supplementary plots). Therefore,
the average level and cyclicality of margins are correlated positively in
Australia but negatively in Germany.

\paragraph{3. How Can Cycles Be Less Volatile Than Non-Cycles?}

Cyclicality implies systematic---but not necessarily large---movements; not
all large/frequent movements follow cycles. Many German observations exhibit
high volatility without any discernible patterns, which explains the
existence of \textquotedblleft volatile non-cycles\textquotedblright \ in
the data.

\paragraph{4. Why Existing Methods Find\  \textquotedblleft Positive
Correlations.\textquotedblright}

These methods' threshold rules tend to recognize high-mean, high-volatility
cases as \textquotedblleft cycles\textquotedblright \ because only
sufficiently large movements can satisfy these conditions (see Appendix C.3
for supplementary plots). In Germany, however, volatility is a poor
predictor of cyclicality (see Question 3 above).

\paragraph{5. Could Intra-Day Cycles Be the Source of Curious Patterns in
Germany?}

The answer is \textquotedblleft yes\textquotedblright \ and
\textquotedblleft no.\textquotedblright \ In general, our daily sampling
frequency and 90-day window are suitable for identifying cycles with the
frequencies of several days to a month or so. Shorter frequencies may not be
well represented.

Nevertheless, if the \textquotedblleft intra-day\textquotedblright \ cycles
follow the frequency of exactly 24 hours (or any hours that can divide 24
evenly), they would be \textquotedblleft averaged out\textquotedblright \ in
the process of computing daily prices\ and would not affect our
observations. The existing studies suggest that they do follow exactly
24-hour cycles (see section 3.3). Hence, how intra-day cycles affect the
multi-daily volatility in our data is not obvious.\footnote{%
One possibility is the existence of \textquotedblleft medium
frequency\textquotedblright \ cycles that are longer than 24 hours, but
shorter than 3--4 days. However, we are not aware of any studies that
document such cycles. In short, the coexistence of daily, weekly, and other
cycles and their interactions constitute an open-ended question for further
research.}

\paragraph{6. Why Manual Classification Provides a Relevant Benchmark.}

At this point, one might question (again) the relevance of human recognition
as a benchmark. Our answer is still the same as in section 1 (paragraph 6):
It is the \textquotedblleft second best\textquotedblright \ option. If we
had a perfect mathematical definition, no detection problem would arise in
the first place. In the absence of such a formula, the existing research
relied on rules of thumb that were ultimately validated by selective
eyeballing by the authors. We made this process more systematic and
transparent.

Even if we had a perfect definition, that would not stop consumers and
politicians from eyeballing and complaining. In the end, their cognitions
and reactions play an important role in determining whether Edgeworth cycles
become a public-policy issue. For these reasons, we believe a
\textquotedblleft ground truth\textquotedblright \ based on the consensus of
reasonably well-educated student RAs provides a relevant benchmark.%
\clearpage

\section{Practical Recommendations}

Based on our findings in sections 5--7, we suggest the following steps as a
practical (but not necessarily the most rigorous) guide for automating the
detection of Edgeworth cycles:

\begin{enumerate}
\item Choose the data frequency and time window that would permit the
identification of hypothesized cycles. That is, the sampling frequency must
be shorter than that of suspected cycles, and the time horizon should
accommodate at least a few repetitions. (For the sake of simple exposition,
our explanation in the following keeps assuming the daily frequency and the
quarterly window.)

\item Eyeball and manually categorize a random sample of 100 station-quarter
observations in terms of cyclicality (but not necessarily asymmetry). If
sufficient numbers of both cyclical and non-cyclical cases are found,
proceed to the next step. If not, increase the sample size.

\item As a first attempt to algorithmically distinguish cycles from
non-cycles, calibrate one of the simpler methods. We recommend the
two-parameter model of Method 4 (MBPI) because it is the only one (among
Methods 1--4) that captures the notion of cyclicality.

\item For more formal, mathematical definitions of cyclicality, use Methods
5 (FT) or 6 (LS), both of which are readily implementable in many
programming languages for scientific computing. Method 7\ (CS) is another
option with similarly off-the-shelf\ implementations.

\item If the performance of these methods is unsatisfactory, try Methods 9
(E-RF), 8 (LSTM), and 10 (E-LSTM), in increasing order of complexity and
expected accuracy.

\item Once the detection of cyclicality (as recognized by humans) is
successfully automated, refine the classification of \textquotedblleft
cycling\textquotedblright \ observations in terms of asymmetry. The
median-price-change statistic from Method 3 (NMC) offers a simple way to
capture asymmetry. For example, one can distinguish between the
Edgeworth-type asymmetry (i.e., the median change is negative), the
inverse-Edgeworth asymmetry (i.e., the median change is positive), and
symmetry (i.e., the median change is approximately zero). Methods 1 (PRNR)
and 2 (MIMD) can be used for the same purpose.

\item If desired, this asymmetry-based classification can be automated by
using some clustering algorithm on the distribution (e.g., a histogram) of
the median price change across station-quarter observations. This process
can be designed as either supervised or unsupervised machine-learning tasks.

\item Once the classification based on both cyclicality and asymmetry is
complete, compute the mean margin and other statistics for each type of
observation (e.g., Table \ref{Table - Margins and cycles}). Welch's $t$
statistic and the associated degrees of freedom can be used for testing the
null hypothesis that the means of the two subsamples (of potentially
different sizes) are equal.

\item The previous step assumes that the dataset contains only prices and
margins. If additional data are available on the characteristics of gasoline
stations and their locations (as well as other demand- and supply-side
factors such as competition), control for these additional covariates in a
suitable regression model.

\item At any point after step 4, one might also consider another refinement
based on the frequency of cycles. Cycles of multiple lengths may coexist
within a single dataset (see sections 3.3 and 7.2). Methods 5--7 would be
useful for this purpose.
\end{enumerate}

Thus, even though Method 10 (E-LSTM) is the top runner in terms of
cycle-detection accuracy, other methods (including the existing ones) have
important roles to play, both as a tool for initial inspection and as a
summary statistic for refinement.

\section{Conclusion}

We propose scalable methods to detect Edgeworth cycles so that the growing
amount of \textquotedblleft big data\textquotedblright \ on fuel prices can
be scrutinized. The failure of the existing methods in noisy data suggests
further investigation would benefit from distinguishing \textquotedblleft
cyclicality\textquotedblright \ from \textquotedblleft
asymmetry.\textquotedblright \ Our nonparametric methods achieve the highest
accuracy; such flexible models typically require large amounts of training
data, but the requirement is minimal in this context. Whether researchers
discover a positive or negative statistical relationship between markups and
cycles depends on the choice of method. Because such \textquotedblleft
facts\textquotedblright \ are supposed to inform regulations and competition
policy, these methodological considerations are directly policy relevant.%
\clearpage

\section*{Appendix A \ Methodological Details}

\subsection*{A.1 \ Details of the New Methods}

\paragraph{Fourier Transform (Method 5).}

The Fourier transform of a continuous function $g\left( x\right) $ is%
\begin{equation}
G\left( f\right) \equiv \int_{-\infty }^{\infty }g\left( x\right) e^{-2\pi
ifx}dx.  \label{eq - Fourier (continuous)}
\end{equation}%
Let us define the Fourier transform operator $\mathcal{F}$ such that $%
\mathcal{F}\left \{ g\right \} =G$, which is a linear operation. A
sinusoidal signal (i.e., sine wave) with frequency $f_{0}$ has a Fourier
transform consisting of a weighted sum of the Dirac delta functions at $\pm
f_{0}$.\footnote{%
The Dirac delta function is $\delta \left( f\right) \equiv \int_{-\infty
}^{\infty }e^{-2\pi ifx}dx$, and hence, we can write $\mathcal{F}\left \{
e^{2\pi f_{0}x}\right \} =\delta \left( f-f_{0}\right) $. The linearity of $%
\mathcal{F}$ and Euler's formula for the complex exponential ($e^{ix}=\cos
x+i\sin x$) lead to the following identities: $\mathcal{F}\left \{ \cos
\left( 2\pi f_{0}x\right) \right \} =\frac{1}{2}\left[ \delta \left(
f-f_{0}\right) +\delta \left( f+f_{0}\right) \right] $ and $\mathcal{F}%
\left
\{ \sin \left( 2\pi f_{0}x\right) \right \} =\frac{1}{2i}\left[
\delta \left( f-f_{0}\right) +\delta \left( f+f_{0}\right) \right] $. See
VanderPlas (2018) for further details.} The practical implication of these
properties is that any signal made up of a sum of sinusoidal components will
have a Fourier transform consisting of a sum of delta functions that mark
the frequencies of those sinusoids. Thus, the Fourier transform directly
measures additive periodic content in a continuous function. The power
spectral density (PSD, or the power spectrum) of a function,%
\begin{equation}
P_{g}\equiv \left \vert \mathcal{F}\left \{ g\right \} \right \vert ^{2},
\label{eq - power spectrum}
\end{equation}%
is a positive, real-valued function of frequency $f$, and provides a
convenient way to quantify the contribution of each frequency $f$ to the
signal $g\left( x\right) $.

When a continuous time series is sampled at regular time intervals with
spacing $\Delta x$, as is the case in our data, one can use the discrete
version of (\ref{eq - Fourier (continuous)}):%
\begin{equation}
G_{obs}\left( f\right) =\sum_{n=-\infty }^{\infty }g\left( n\Delta x\right)
e^{-2\pi ifn\Delta x}.  \label{eq - Fourier (discrete)}
\end{equation}%
Acknowledging the finite sample size $N$ and focusing on the relevant
frequency range $0\leq f\leq \frac{1}{\Delta x}$, one can define $N$ evenly
spaced frequencies with $\Delta f=\frac{1}{N\Delta x}$ covering this range.
Let $g_{n}\equiv g\left( n\Delta x\right) $ and $G_{k}\equiv G_{obs}\left(
k\Delta f\right) $. Then, the sample analog of (\ref{eq - Fourier
(continuous)}) is%
\begin{equation}
G_{k}=\sum_{n=0}^{N}g_{n}e^{-2\pi ikn/N}.
\label{eq - Fourier (sample analog)}
\end{equation}%
One can construct the sample analog of the Fourier power spectrum (\ref{eq -
power spectrum}) as (\ref{eq - classical periodogram}) in the main text.
This is the \textquotedblleft classical\textquotedblright \ or
\textquotedblleft Schuster\textquotedblright \ periodogram.\footnote{%
See Press et al. (1992, section 12.2) for computational implementation.}

\bigskip

A potential drawback of the threshold rule in (\ref{eq - condition A
(Fourier, max)}) is that it exclusively focuses on the highest point and
ignores the rest. As an alternative rule, we can compare the highest point
with the heights of other, less powerful frequencies. One way to capture
relative heights of multiple frequencies is to measure the \textquotedblleft
concentration\textquotedblright \ of power in a limited number of
frequencies. We use the\ Herfindahl-Hirschman Index (HHI) for an additional
check for \textquotedblleft significant\textquotedblright \ cycles:%
\begin{equation}
HHI_{i,t}\equiv \sum_{f}\left( \frac{P_{i,t}\left( f\right) }{%
\sum_{f}P_{i,t}\left( f\right) }\right) ^{2}>\theta _{hhi}^{FT},
\label{eq - condition B (Fourier, HHI)}
\end{equation}%
where $\theta _{hhi}^{FT}\in \left( 0,1\right] $\ is a scalar threshold
parameter.\footnote{%
The HHI is a summary statistic that is typically used to measure the degree
of market-share concentration in oligopolistic industries. A high value of
the HHI indicates the market is close to monopoly.} A high value of $%
HHI_{i,t}$ indicates strong periodicity at certain frequencies relative to
other, weaker frequencies.

\paragraph{Lomb-Scargle Periodogram (Method 6).}

Even though the classical periodogram in (\ref{eq - classical periodogram})
appears different from (\ref{eq - Lomb-Scargle periodogram}), (\ref{eq -
classical periodogram}) can be rewritten as%
\begin{equation*}
P\left( f\right) =\frac{1}{N}\left[ \left( \sum_{n}g_{n}\cos \left( 2\pi
fx_{n}\right) \right) ^{2}+\left( \sum_{n}g_{n}\sin \left( 2\pi
fx_{n}\right) \right) ^{2}\right] .
\end{equation*}%
Thus, the only major difference between (\ref{eq - classical periodogram})
and (\ref{eq - Lomb-Scargle periodogram}) is the denominators in (\ref{eq -
Lomb-Scargle periodogram}).

Statistically, one can interpret the Lomb-Scargle periodogram as a
collection of least-squares regressions in which one fits a sinusoidal model
at each frequency $f$:%
\begin{equation}
\hat{g}\left( x;f\right) =A_{f}\sin \left( 2\pi f\left( x-\phi _{f}\right)
\right) ,  \label{eq - sinusoidal model at each frequency}
\end{equation}%
where amplitude $A_{f}$ and phase $\phi _{f}$ are the parameters to be
estimated by minimizing the sum of squared residuals:%
\begin{equation}
SSR^{LS}\left( f\right) \equiv \sum_{n}\left( g_{n}-\hat{g}\left(
x_{n};f\right) \right) ^{2}.  \label{eq - SSR (Lomb-Scargle)}
\end{equation}%
Scargle (1982) shows the following periodogram is identical to (\ref{eq -
Lomb-Scargle periodogram}):%
\begin{equation*}
\tilde{P}^{LS}\left( f\right) =\frac{1}{2}\left[ SSR_{0}^{LS}-SSR^{LS}\left(
f\right) \right] ,
\end{equation*}%
where $SSR_{0}^{LS}$ is the sum of squared residuals from the restricted
model in which the only regressor is a constant term. The idea is that the
frequencies with good fit will exhibit high $\tilde{P}^{LS}\left( f\right) $.

The HHI variant of the LS method is%
\begin{equation}
HHI_{i,t}^{LS}\equiv \sum_{f}\left( \frac{P_{i,t}^{LS}\left( f\right) }{%
\sum_{f}P_{i,t}^{LS}\left( f\right) }\right) ^{2}>\theta _{hhi}^{LS}.
\label{eq - condition B (Lomb-Scargle, HHI)}
\end{equation}

\paragraph{Cubic Splines (Method 7).}

A spline is a piecewise polynomial function:%
\begin{equation}
S_{K}\left( x\right) =\sum_{j=0}^{P}\beta _{j}x^{j}+\sum_{k=1}^{N}\beta
_{P+k}\left( x-\tau _{k}\right) ^{P}\mathbb{I}\left \{ x\geq \tau _{k}\right
\} ,  \label{eq - spline}
\end{equation}%
where $K=1+P+N$ is the number of coefficients, $P$ is the order of the
polynomial (not to be confused with the periodogram in Methods 5--6 or our
notation for the price, $p$), and the support for $x$ is covered by $N+1$
ordered subintervals that are joined by $N$ knots ($\tau _{1}<\tau
_{2}<\cdots <\tau _{N}$).\footnote{%
This $N$ should not be confused with our notation for sample size in the
discrete Fourier transform.} It is a special case of a sieve/series
approximation that constitutes a class of nonparametric regression methods.%
\footnote{%
Any continuous function can be uniformly well approximated by a polynomial
of sufficiently high order, and the rate of approximation is $o\left(
K^{-2}\right) $. Other series models include trigonometric polynomials,
wavelets, orthogonal wavelets, B-splines, and artificial neural networks.
See Hansen (2020, ch. 20) for an introduction and Chen (2007) for a review.}
We use splines as an interpolator to smooth the discrete (daily) time series
and facilitate further calculations. Specifically, we use a cubic Hermite
interpolator, which is a spline where each piece is a third-degree
polynomial of Hermite form (i.e., $P=3$, $N=88$, and $\beta $s are
prespecified).\footnote{%
On the unit interval $d\in \left( 0,1\right) $, given a starting point $%
p_{0} $ at $d=0$, an ending point $p_{1}$ at $d=1$, and slopes $m_{0}$ and $%
m_{1}$, this polynomial is%
\begin{equation*}
p\left( d\right) =\left( 2d^{3}-3d^{2}+1\right) p_{0}+\left(
d^{3}-2d^{2}+d\right) m_{0}+\left( -2d^{3}+3d^{2}\right) p_{1}+\left(
d^{3}-d^{2}\right) m_{1}.
\end{equation*}%
This form ensures the observed values $\left( p_{0},p_{1}\right) $ and their
slopes $\left( m_{0},m_{1}\right) $ are fitted exactly. It has become a
default specification of CS in SciPy, a set of commonly used Python
libraries for scientific computing.}

In addition to the indicator of frequent oscillations in (\ref{eq -
condition (CS roots)}), we propose a measure that captures amplitude as
well. We subtract the lowest daily price in $\left( i,t\right) $ from all of
its daily prices, $\underline{p}_{i,d}\equiv p_{i,d}-min_{d\in t}\left(
p_{i,d}\right) $, fit CS to $\left( \underline{p}_{i,d}\right) _{d\in t}$,
and calculate its integral over $d\in \left[ 1,90\right] $. We set $%
cycle_{i,t}=1$ if and only if%
\begin{equation}
\int_{1}^{90}\underline{CS}_{i,t}\left( d\right) >\theta _{int}^{CS},
\label{eq - condition (CS integral)}
\end{equation}%
where $\underline{CS}_{i,t}\left( d\right) $ is the fitted value of $%
\underline{p}_{i,d}$ at time $d$. Because this definite integral equals the
area between the price series and its lowest level within $\left( i,t\right) 
$, this condition captures cycles with large amplitude and sustained high
prices.

We also construct a discrete (raw data) analog of the splines-integral
measure as follows:%
\begin{equation}
\sum_{d=1}^{90}\left \vert \bar{p}_{i,d}\right \vert >\theta _{abs}^{CS},
\label{eq - condition (CS absolute)}
\end{equation}%
where $\bar{p}_{i,d}$ is the demeaned price. The information content of this
statistic is similar to the previous one, but its calaculation is simpler.

\paragraph{Long Short-Term Memory (Method 8).}

Compared with Greff et al.'s (2017) \textquotedblleft
vanilla\textquotedblright \ setup, we make two simplifications. First, our
law of motion for $\mathbf{c}_{d}^{l}$\ (\ref{eq - multi-layer LSTM state 2}%
) uses the same set of parameters $\left( \mathbf{\omega }_{7}^{l},\mathbf{%
\omega }_{8}^{l},\mathbf{\omega }_{9}^{l}\right) $ twice. This
simplification corresponds to their \textquotedblleft Coupled Input and
Forget Gate\textquotedblright \ variant due to Cho et al. (2014), which is
also referred to as Gated Recurrent Units (GRUs) in the literature. Second,
we do not include $\mathbf{c}_{d}^{l}$\ or $\mathbf{c}_{d-1}^{l}$\ inside $%
\Lambda $ in (\ref{eq - multi-layer LSTM state 1}) or inside $\tanh $ and $%
\Lambda $\ in (\ref{eq - multi-layer LSTM state 2}). This omission
corresponds to their \textquotedblleft No Peepholes\textquotedblright \
variant. Greff et al. (2017) show these simplifications reduce the number of
parameters without compromising predictive accuracy.

We implement LSTM in TensorFlow-GPU 2.6 (tf.keras.models.Sequential). Our
choice of network architecture and activation functions---which constitute
the specification of effective functional forms---are as explained in the
main text. The total number of weight parameters is 2,165. We set other
tuning parameters and the details of numerical optimization as follows: (i)
the dropout rate is 0.5, (ii) the optimizer is tf.keras.optimizer.RMSprop
with the learning rate of 0.0005, (iii) the number of epochs is 100, and
(iv) the batch size is 30.

\paragraph{Ensemble in Random Forests (Method 9).}

The relationship between \textquotedblleft decision trees\textquotedblright
\ and \textquotedblleft random forests\textquotedblright \ is as follows,
according to Murphy (2012, ch. 16). Because finding the truly optimal
partitioning in a decision-trees model is computationally infeasible, some
greedy, iterative procedures are used in the estimation/tuning of the
parameters $\left( \mathbf{\omega }^{RF},\mathbf{\kappa }^{RF}\right) $.
However, the hierarchical nature of this process leads to unstable
predictions. Averaging over multiple estimates from bootstrapped subsamples
(\textquotedblleft bootstrap aggregating\textquotedblright \ or
\textquotedblleft bagging\textquotedblright ) is a commonly used technique
to reduce this variance. A further improvement is possible by randomly
choosing a subset of input variables, in addition to \textquotedblleft
bagging.\textquotedblright \ This technique is called \textquotedblleft
random forests\textquotedblright \ (Breiman 2001a) and is known to perform
well in many different contexts (e.g., Caruana and Niculescu-Mizil 2006).

We implement E-RF in scikit-learn 0.24.2
(sklearn.ensemble.RandomForestClassifier), with default options for all
settings.

\paragraph{Ensemble in Long Short-Term Memory (Method 10).}

Our E-LSTM implementation details are the same as in the basic LSTM (Method
8). The only difference is that the total number of weight parameters is
larger at 2,933 to incorporate the additional input variables from Methods
1--7.

\subsection*{A.2 \ Parameter Optimization}

We define two types of prediction errors as follows:%
\begin{eqnarray}
\% \text{ false negative}\left( \mathbf{\theta }\right) &\equiv &\frac{%
\sum_{\left( i,t\right) }\mathbb{I}\left \{ \widehat{cycle}_{i,t}\left( 
\mathbf{\theta }\right) =0,cycle_{i,t}=1\right \} }{\# \text{ }all\text{ }%
predictions}\times 100,\text{ and}  \label{eq - percent false negative} \\
\% \text{ false positive}\left( \mathbf{\theta }\right) &\equiv &\frac{%
\sum_{\left( i,t\right) }\mathbb{I}\left \{ \widehat{cycle}_{i,t}\left( 
\mathbf{\theta }\right) =1,cycle_{i,t}=0\right \} }{\# \text{ }all\text{ }%
predictions}\times 100.  \label{eq - percent false positive}
\end{eqnarray}%
They correspond to type II errors and type I errors in statistics,
respectively.

We occasionally encounter cases in which a range of parameter values attain
the same (maximum) accuracy. In such cases, we report the median of all $%
\mathbf{\theta }^{\ast }$\ values that we find in our grid search. These
cases typically involve \textquotedblleft degenerate\textquotedblright \
predictions in which $\widehat{cycle}_{i,t}\left( \mathbf{\theta }\right) =1$
or $\widehat{cycle}_{i,t}\left( \mathbf{\theta }\right) =0$ for all $\left(
i,t\right) $, and hence are mostly irrelevant for the purpose of finding
well-performing $\mathbf{\theta }$s.\clearpage

\section*{Appendix B \ Additional Results}

\subsection*{B.1 \ Variants of FT, LS, and CS}

Table \ref{Table - Accuracy (additional results)} reports the performances
of the variants of Methods 5, 6, and 7. In Methods 5 and 6, the
\textquotedblleft max\textquotedblright \ and \textquotedblleft
HHI\textquotedblright \ variants are as explained in section 4.2 and
Appendix A.1. The \textquotedblleft peak\textquotedblright \ variant is
similar to the \textquotedblleft max\textquotedblright \ one except that we
additionally use a peak-detection algorithm to ensure we are measuring the
height of the highest (and well-behaved) peak in the power spectrum and not
some accidental maximum due to noisy data. In Method 7, the
\textquotedblleft roots\textquotedblright \ variant is the baseline version
in section 4.2. Its \textquotedblleft integral\textquotedblright \ and
\textquotedblleft absolute value\textquotedblright \ variants are explained
in Appendix A.1.

\subsection*{B.2 \ Data Requirement and Marginal Cost of Accuracty}

Table \ref{Table - Return on data} reports the means of accuracy (\%
correct) across 101 bootstrap sample splits that are underlying the visual
summaries in Figure \ref{Figure - Gains from Data} in section 6. Table \ref%
{Table - Return on data (stdev)} shows the standard deviations of accuracy
are usually less than 1 percentage point when more than 1\% of the sample is
used for training.

\begin{figure}[htb!!!!]\centering%
\caption{Marginal Costs of Accuracy (E-LSTM)}%
\includegraphics[width=0.90\textwidth]{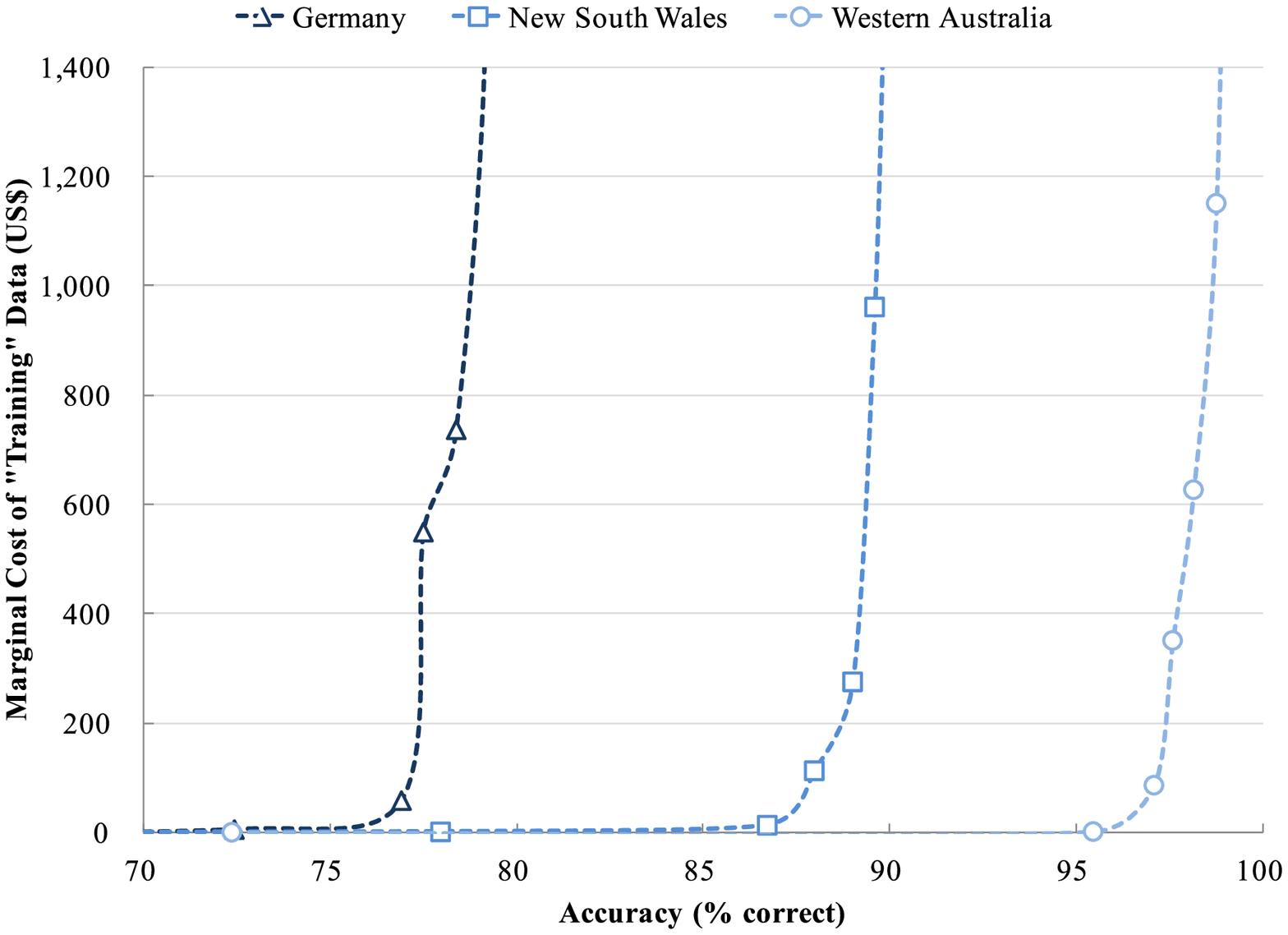}
\caption*{\footnotesize {\textit{Note}: These marginal-cost curves
are based on the numbers reported in Table \ref{Table - Return on data}
(shown with markers) and a splines-based interpolation (dashed lines).}}%
\label{Figure - Marginal Costs}
\end{figure}

Figure \ref{Figure - Marginal Costs} and Panel (C) of Table {\footnotesize {%
\ref{Table - Return on data}}} show the \textquotedblleft marginal costs of
accuracy\textquotedblright \ (i.e., the amount of RA work required for an
extra percentage-point increase in accuracy). The marginal cost is initially
low with only a few cents, but rapidly increases as we approach the maximum
possible accuracy levels. Because the difficulty of accurate classification
in a new dataset is unknown \textit{a priori}, one cannot set realistic
targets without some preliminary analysis. Nevertheless, our findings in
section 5 are encouraging in that only a few hundred labeled observations
are necessary to reach approximately optimal accuracy levels.

\begin{table}[tbh]
\caption{Performance of Automatic Detection Methods (Other Variants)}
\label{Table - Accuracy (additional results)}
\begin{center}
\fontsize{9pt}{11pt}\selectfont%
\begin{tabular}{lccccccccc}
\hline \hline
& $\left( 5\right) $ & $(5^{\prime })$ & $(5^{\prime \prime })$ & $\left(
6\right) $ & $(6^{\prime })$ & $(6^{\prime \prime })$ & $\left( 7\right) $ & 
$\left( 7^{\prime }\right) $ & $(7^{\prime \prime })$ \\ 
Method & FT$_{\max }$ & FT$_{peak}$ & FT2$_{hhi}$ & LS$_{\max }$ & LS$%
_{peak} $ & LS$_{hhi}$ & CS$_{roots}$ & CS$_{int}$ & CS$_{abs}$ \\ \hline
\multicolumn{1}{c}{} &  &  &  &  &  &  &  &  &  \\ 
\multicolumn{10}{c}{\textit{I. Western Australia} (\# manually labeled
observations: $24,569$)} \\ 
\multicolumn{1}{c}{} &  &  &  &  &  &  &  &  &  \\ 
Parameter 1 & $0.12$ & $0.14$ & $0.04$ & $0.21$ & $0.23$ & $0.44$ & $22.50$
& $551.47$ & $246.08$ \\ 
Parameter 2 & $-$ & $-$ & $-$ & $-$ & $-$ & $-$ & $-$ & $-$ & $-$ \\ 
\% correct (median) & $90.11$ & $88.40$ & $87.61$ & $90.15$ & $89.66$ & $%
81.83$ & $85.47$ & $83.42$ & $85.14$ \\ 
(Standard deviations) & $\left( 0.40\right) $ & $\left( 0.45\right) $ & $%
\left( 0.39\right) $ & $\left( 0.36\right) $ & $\left( 0.43\right) $ & $%
\left( 0.54\right) $ & $\left( 0.45\right) $ & $\left( 0.54\right) $ & $%
\left( 0.42\right) $ \\ 
of which cycling & $58.24$ & $57.31$ & $59.12$ & $57.92$ & $57.10$ & $54.13$
& $56.41$ & $55.92$ & $57.14$ \\ 
of which not & $31.87$ & $31.09$ & $28.49$ & $32.23$ & $32.56$ & $27.70$ & $%
29.06$ & $27.49$ & $28.00$ \\ 
\% false negative & $2.48$ & $4.05$ & $1.91$ & $3.30$ & $4.50$ & $6.15$ & $%
5.29$ & $4.82$ & $3.32$ \\ 
\% false positive & $7.41$ & $7.5$ & $10.48$ & $6.55$ & $5.84$ & $12.03$ & $%
9.24$ & $11.76$ & $11.54$ \\ 
\multicolumn{1}{c}{} &  &  &  &  &  &  &  &  &  \\ 
\multicolumn{10}{c}{\textit{II. New South Wales} (\# manually labeled
observations: $9,693$)} \\ 
\multicolumn{1}{c}{} &  &  &  &  &  &  &  &  &  \\ 
Parameter 1 & $0.20$ & $0.27$ & $0.21$ & $0.57$ & $0.81$ & $29.21$ & $4.50$
& $783.11$ & $459.83$ \\ 
Parameter 2 & $-$ & $-$ & $-$ & $-$ & $-$ & $-$ & $-$ & $-$ & $-$ \\ 
\% correct (median) & $80.71$ & $81.85$ & $81.23$ & $80.82$ & $82.21$ & $%
81.38$ & $73.90$ & $75.45$ & $79.63$ \\ 
(Standard deviations) & $\left( 0.80\right) $ & $\left( 0.70\right) $ & $%
\left( 0.83\right) $ & $\left( 0.80\right) $ & $\left( 0.81\right) $ & $%
\left( 0.84\right) $ & $\left( 0.89\right) $ & $\left( 0.79\right) $ & $%
\left( 0.87\right) $ \\ 
of which cycling & $66.53$ & $66.99$ & $64.72$ & $66.43$ & $66.89$ & $67.30$
& $70.40$ & $68.13$ & $67.25$ \\ 
of which not & $14.18$ & $14.85$ & $16.50$ & $14.39$ & $15.32$ & $14.08$ & $%
3.51$ & $7.32$ & $12.38$ \\ 
\% false negative & $5.47$ & $4.54$ & $6.19$ & $4.02$ & $4.07$ & $4.54$ & $%
0.77$ & $3.20$ & $3.56$ \\ 
\% false positive & $13.82$ & $13.62$ & $12.58$ & $15.16$ & $13.72$ & $14.08$
& $25.32$ & $21.35$ & $16.81$ \\ 
\multicolumn{1}{c}{} &  &  &  &  &  &  &  &  &  \\ 
\multicolumn{10}{c}{\textit{III. Germany} (\# manually labeled observations: 
$35,685$)} \\ 
\multicolumn{1}{c}{} &  &  &  &  &  &  &  &  &  \\ 
Parameter 1 & $0.24$ & $0.90$ & $0.67$ & $0.62$ & $1.93$ & $42.96$ & $24.50$
& $994.19$ & $4,623$ \\ 
Parameter 2 & $-$ & $-$ & $-$ & $-$ & $-$ & $-$ & $-$ & $-$ & $-$ \\ 
\% correct (median) & $60.50$ & $60.57$ & $60.35$ & $60.36$ & $60.50$ & $%
60.52$ & $71.28$ & $60.29$ & $60.49$ \\ 
(Standard deviations) & $\left( 0.56\right) $ & $\left( 0.50\right) $ & $%
\left( 0.53\right) $ & $\left( 0.59\right) $ & $\left( 0.57\right) $ & $%
\left( 0.53\right) $ & $\left( 0.42\right) $ & $\left( 0.48\right) $ & $%
\left( 0.48\right) $ \\ 
of which cycling & $0.00$ & $0.00$ & $0.00$ & $0.00$ & $0.00$ & $24.30$ & $%
25.88$ & $0.00$ & $0.00$ \\ 
of which not & $60.50$ & $60.57$ & $60.35$ & $60.36$ & $60.50$ & $47.00$ & $%
45.40$ & $60.29$ & $60.49$ \\ 
\% false negative & $39.50$ & $39.41$ & $39.65$ & $39.57$ & $39.50$ & $15.26$
& $14.28$ & $39.51$ & $0.00$ \\ 
\% false positive & $0.00$ & $0.01$ & $0.00$ & $0.07$ & $0.00$ & $13.43$ & $%
14.45$ & $0.20$ & $39.51$ \\ \hline \hline
\end{tabular}
\begin{minipage}{475pt}
{\fontsize{9pt}{9pt}\selectfont \smallskip  \textit{Note}: See the text of Appendix sections A.1 and B.1 for the definition of each method.}
\end{minipage}
\end{center}
\end{table}

\begin{table}[tbh]
\caption{Benefits and Costs of Additional Data}
\label{Table - Return on data}
\begin{center}
\fontsize{9pt}{11pt}\selectfont%
\begin{tabular}{lcccccccc}
\hline \hline
& $(1)$ & $(2)$ & $(3)$ & $(4)$ & $\left( 5\right) $ & $\left( 6\right) $ & $%
\left( 7\right) $ & $\left( 8\right) $ \\ 
Subsample used for \textquotedblleft training\textquotedblright & $0.1\%$ & $%
1\%$ & $5\%$ & $10\%$ & $20\%$ & $40\%$ & $60\%$ & $80\%$ \\ \hline
\multicolumn{9}{l}{(A) Median Accuracy (\% correct)} \\ 
\multicolumn{1}{c}{} &  &  &  &  &  &  &  &  \\ 
\multicolumn{9}{c}{\textit{I. Western Australia} (\# manually labeled
observations: $24,569$)} \\ 
1. PRNR & $89.88$ & $90.64$ & $90.70$ & $90.73$ & $90.72$ & $90.80$ & $90.80$
& $90.80$ \\ 
2. MIMD & $90.55$ & $91.21$ & $91.29$ & $91.28$ & $91.31$ & $91.29$ & $91.29$
& $91.27$ \\ 
3. NMC & $89.33$ & $89.35$ & $89.34$ & $89.36$ & $89.36$ & $89.36$ & $89.35$
& $89.34$ \\ 
4. MBPI & $89.95$ & $90.15$ & $90.24$ & $90.27$ & $90.26$ & $90.25$ & $90.26$
& $90.23$ \\ 
5. FT & $89.29$ & $89.98$ & $90.06$ & $90.06$ & $90.08$ & $90.12$ & $90.15$
& $90.11$ \\ 
6. LS & $89.67$ & $89.92$ & $90.01$ & $90.01$ & $90.03$ & $90.06$ & $90.06$
& $90.15$ \\ 
7. CS & $84.22$ & $84.96$ & $85.48$ & $85.48$ & $85.51$ & $85.52$ & $85.43$
& $85.47$ \\ 
8. LSTM & $85.32$ & $95.54$ & $97.11$ & $97.66$ & $98.20$ & $98.78$ & $99.07$
& $99.25$ \\ 
9. E-RF & $90.78$ & $96.53$ & $97.86$ & $98.27$ & $98.59$ & $98.87$ & $98.96$
& $99.04$ \\ 
10. E-LSTM & $72.36$ & $95.46$ & $97.09$ & $97.59$ & $98.15$ & $98.76$ & $%
99.06$ & $99.25$ \\ 
\multicolumn{1}{c}{} &  &  &  &  &  &  &  &  \\ 
\multicolumn{9}{c}{\textit{II. New South Wales} (\# manually labeled
observations: $9,693$)} \\ 
1. PRNR & $75.78$ & $78.12$ & $78.51$ & $78.50$ & $78.54$ & $78.49$ & $78.49$
& $78.55$ \\ 
2. MIMD & $77.81$ & $78.13$ & $78.24$ & $78.29$ & $78.31$ & $78.42$ & $78.29$
& $78.39$ \\ 
3. NMC & $70.95$ & $70.95$ & $70.95$ & $70.93$ & $70.94$ & $70.99$ & $70.94$
& $70.96$ \\ 
4. MBPI & $80.47$ & $80.74$ & $81.25$ & $81.27$ & $81.35$ & $81.40$ & $81.48$
& $81.59$ \\ 
5. FT & $76.48$ & $79.93$ & $80.56$ & $80.62$ & $80.64$ & $80.57$ & $80.63$
& $80.71$ \\ 
6. LS & $77.17$ & $80.29$ & $80.66$ & $80.67$ & $80.70$ & $80.64$ & $80.58$
& $80.82$ \\ 
7. CS & $72.02$ & $73.80$ & $73.83$ & $73.84$ & $73.89$ & $73.90$ & $73.85$
& $73.90$ \\ 
8. LSTM & $60.60$ & $78.84$ & $86.52$ & $87.64$ & $88.68$ & $89.13$ & $89.45$
& $89.63$ \\ 
9. E-RF & $78.99$ & $83.00$ & $84.40$ & $85.09$ & $85.85$ & $86.59$ & $86.85$
& $87.42$ \\ 
10. E-LSTM & $44.39$ & $77.96$ & $86.72$ & $87.98$ & $89.01$ & $89.60$ & $%
89.94$ & $90.30$ \\ 
\multicolumn{1}{c}{} &  &  &  &  &  &  &  &  \\ 
\multicolumn{9}{c}{\textit{III. Germany} (\# manually labeled observations: $%
35,685$)} \\ 
1. PRNR & $55.48$ & $60.43$ & $60.42$ & $60.46$ & $60.42$ & $60.45$ & $60.47$
& $60.38$ \\ 
2. MIMD & $60.19$ & $60.51$ & $60.58$ & $60.58$ & $60.60$ & $60.66$ & $60.63$
& $60.61$ \\ 
3. NMC & $60.42$ & $60.43$ & $60.44$ & $60.44$ & $60.46$ & $60.47$ & $60.43$
& $60.53$ \\ 
4. MBPI & $64.49$ & $65.16$ & $65.25$ & $65.26$ & $65.32$ & $65.32$ & $65.35$
& $65.39$ \\ 
5. FT & $60.42$ & $60.42$ & $60.42$ & $60.43$ & $60.40$ & $60.46$ & $60.45$
& $60.50$ \\ 
6. LS & $60.42$ & $60.42$ & $60.43$ & $60.44$ & $60.42$ & $60.44$ & $60.46$
& $60.36$ \\ 
7. CS & $70.66$ & $71.26$ & $71.31$ & $71.29$ & $71.29$ & $71.28$ & $71.31$
& $71.28$ \\ 
8. LSTM & $60.45$ & $65.97$ & $70.83$ & $72.40$ & $73.74$ & $74.27$ & $74.43$
& $74.61$ \\ 
9. E-RF & $66.38$ & $71.86$ & $74.78$ & $75.04$ & $75.30$ & $75.79$ & $75.79$
& $76.14$ \\ 
10. E-LSTM & $60.45$ & $72.43$ & $76.91$ & $77.50$ & $78.38$ & $79.21$ & $%
79.51$ & $79.58$ \\ 
\multicolumn{1}{c}{} &  &  &  &  &  &  &  &  \\ 
\multicolumn{9}{l}{(B) Total Costs of Manual Labeling (US\$)} \\ 
I. Western Australia & $3.51$ & $35.1$ & $176$ & $351$ & $702$ & $1,404$ & $%
2,106$ & $2,808$ \\ 
II. New South Wales & $2.84$ & $28.4$ & $142$ & $284$ & $567$ & $1,134$ & $%
1,701$ & $2,268$ \\ 
III. Germany & $6.48$ & $64.8$ & $324$ & $648$ & $1,296$ & $2,592$ & $3,888$
& $5,184$ \\ 
\multicolumn{1}{c}{} &  &  &  &  &  &  &  &  \\ 
\multicolumn{9}{l}{(C) E-LSTM's Marginal Costs of Accuracy (US\$ per correct
\% point)} \\ 
I. Western Australia & $0.05$ & $1.37$ & $86.13$ & $351$ & $627$ & $1,151$ & 
$2,340$ & $3,695$ \\ 
II. New South Wales & $0.06$ & $0.76$ & $12.95$ & $113$ & $275$ & $961$ & $%
1,668$ & $1,575$ \\ 
III. Germany & $0.11$ & $4.87$ & $57.86$ & $549$ & $736$ & $1,561$ & $4,320$
& $18,514$ \\ \hline \hline
\end{tabular}
\begin{minipage}{475pt}
{\fontsize{9pt}{9pt}\selectfont \smallskip  \textit{Note}: The numbers in panel (A) indicate accuracy in the 20\% testing subsample. Some or all of the remaining 80\% subsample are used as a training subsample to optimize the parameters of each model. We randomly split the sample 101 times, tune the parameters as many times, and report their median performances. The dollar costs in Panels (B) and (C) are based on the total RA hours to manually classify cycles and the hourly wage of \$13.50 (see section 3).}
\end{minipage}
\end{center}
\end{table}

\begin{table}[tbh]
\caption{Standard Deviations of Accuracy under Different Sample Sizes}
\label{Table - Return on data (stdev)}
\begin{center}
\fontsize{9pt}{11pt}\selectfont%
\begin{tabular}{lcccccccc}
\hline \hline
& $(1)$ & $(2)$ & $(3)$ & $(4)$ & $\left( 5\right) $ & $\left( 6\right) $ & $%
\left( 7\right) $ & $\left( 8\right) $ \\ 
Subsample used for \textquotedblleft training\textquotedblright & $0.1\%$ & $%
1\%$ & $5\%$ & $10\%$ & $20\%$ & $40\%$ & $60\%$ & $80\%$ \\ \hline
\multicolumn{9}{l}{Standard Deviation of: Accuracy (\% correct)} \\ 
\multicolumn{1}{c}{} &  &  &  &  &  &  &  &  \\ 
\multicolumn{9}{c}{\textit{I. Western Australia} (\# manually labeled
observations: $24,569$)} \\ 
1. PRNR & $1.94$ & $0.64$ & $0.33$ & $0.11$ & $0.12$ & $0.16$ & $0.22$ & $%
0.37$ \\ 
2. MIMD & $2.09$ & $0.6$ & $0.19$ & $0.10$ & $0.11$ & $0.15$ & $0.20$ & $%
0.38 $ \\ 
3. NMC & $3.22$ & $0.39$ & $0.11$ & $0.08$ & $0.09$ & $0.15$ & $0.23$ & $%
0.38 $ \\ 
4. MBPI & $2.34$ & $0.46$ & $0.11$ & $0.09$ & $0.11$ & $0.14$ & $0.22$ & $%
0.36$ \\ 
5. FT & $1.80$ & $0.50$ & $0.25$ & $0.16$ & $0.14$ & $0.16$ & $0.23$ & $0.40$
\\ 
6. LS & $1.39$ & $0.38$ & $0.14$ & $0.15$ & $0.12$ & $0.16$ & $0.24$ & $0.36$
\\ 
7. CS & $2.15$ & $0.59$ & $0.16$ & $0.09$ & $0.12$ & $0.20$ & $0.26$ & $0.45$
\\ 
8. LSTM & $23.81$ & $1.15$ & $0.20$ & $0.21$ & $0.20$ & $0.21$ & $6.01$ & $%
0.18$ \\ 
9. E-RF & $1.65$ & $0.44$ & $0.17$ & $0.15$ & $0.11$ & $0.09$ & $0.09$ & $%
0.15$ \\ 
10. E-LSTM & $23.22$ & $1.32$ & $0.21$ & $0.19$ & $5.91$ & $0.20$ & $0.16$ & 
$0.14$ \\ 
\multicolumn{1}{c}{} &  &  &  &  &  &  &  &  \\ 
\multicolumn{9}{c}{\textit{II. New South Wales} (\# manually labeled
observations: $9,693$)} \\ 
1. PRNR & $4.94$ & $1.03$ & $0.62$ & $0.53$ & $0.28$ & $0.32$ & $0.49$ & $%
0.85$ \\ 
2. MIMD & $1.90$ & $0.98$ & $0.47$ & $0.28$ & $0.30$ & $0.36$ & $0.52$ & $%
0.88$ \\ 
3. NMC & $12.86$ & $0.04$ & $0.10$ & $0.16$ & $0.21$ & $0.36$ & $0.56$ & $%
0.97$ \\ 
4. MBPI & $2.44$ & $1.51$ & $0.31$ & $0.31$ & $0.25$ & $0.37$ & $0.55$ & $%
0.86$ \\ 
5. FT & $6.57$ & $1.65$ & $0.54$ & $0.36$ & $0.30$ & $0.40$ & $0.48$ & $0.80$
\\ 
6. LS & $4.46$ & $1.10$ & $0.51$ & $0.34$ & $0.27$ & $0.31$ & $0.48$ & $0.80$
\\ 
7. CS & $14.34$ & $1.01$ & $0.43$ & $0.17$ & $0.27$ & $0.41$ & $0.61$ & $%
0.89 $ \\ 
8. LSTM & $20.66$ & $2.57$ & $0.74$ & $0.46$ & $0.29$ & $0.33$ & $0.41$ & $%
0.67$ \\ 
9. E-RF & $6.34$ & $0.94$ & $0.42$ & $0.36$ & $0.34$ & $0.37$ & $0.46$ & $%
0.69$ \\ 
10. E-LSTM & $19.25$ & $5.21$ & $0.89$ & $0.45$ & $0.34$ & $0.34$ & $0.42$ & 
$0.67$ \\ 
\multicolumn{1}{c}{} &  &  &  &  &  &  &  &  \\ 
\multicolumn{9}{c}{\textit{III. Germany} (\# manually labeled observations: $%
35,685$)} \\ 
1. PRNR & $4.35$ & $2.06$ & $0.11$ & $0.08$ & $0.12$ & $0.21$ & $0.31$ & $%
0.49$ \\ 
2. MIMD & $5.30$ & $0.75$ & $0.20$ & $0.28$ & $0.14$ & $0.19$ & $0.32$ & $%
0.50$ \\ 
3. NMC & $5.67$ & $0.38$ & $0.10$ & $0.09$ & $0.14$ & $0.19$ & $0.33$ & $%
0.52 $ \\ 
4. MBPI & $1.64$ & $0.84$ & $0.27$ & $0.19$ & $0.16$ & $0.22$ & $0.28$ & $%
0.52$ \\ 
5. FT & $5.71$ & $0.23$ & $0.11$ & $0.09$ & $0.14$ & $0.20$ & $0.30$ & $0.56$
\\ 
6. LS & $6.97$ & $0.15$ & $0.07$ & $0.10$ & $0.15$ & $0.25$ & $0.33$ & $0.59$
\\ 
7. CS & $2.98$ & $0.70$ & $0.20$ & $0.10$ & $0.13$ & $0.17$ & $0.29$ & $0.42$
\\ 
8. LSTM & $0.68$ & $1.47$ & $1.05$ & $1.08$ & $0.67$ & $0.40$ & $0.38$ & $%
0.44$ \\ 
9. E-RF & $2.99$ & $2.29$ & $2.26$ & $1.33$ & $1.47$ & $1.29$ & $1.47$ & $%
1.46$ \\ 
10. E-LSTM & $0.80$ & $2.08$ & $0.62$ & $0.47$ & $0.63$ & $0.51$ & $0.41$ & $%
0.53$ \\ \hline \hline
\end{tabular}
\begin{minipage}{475pt}
{\fontsize{9pt}{9pt}\selectfont \smallskip  \textit{Note}: The numbers indicate the standard deviations of accuracy across the 101 bootstrap sample-splits.}
\end{minipage}
\end{center}
\end{table}
\clearpage

\clearpage

\subsection*{B.3 \ Using Only \textquotedblleft Cleaner\textquotedblright \
Subsamples}

Table \ref{Table - Accuracy (cleaner data)} and Figure \ref{Figure - Gains
from Data (Cleaner Data)} report alternative results based only on
subsamples that are either unanimously labeled as \textquotedblleft
cycling\textquotedblright \ by all RAs or not labeled as \textquotedblleft
cycling\textquotedblright \ by any RA, thereby ignoring ambiguous
observations. Accuracy is higher overall, but relative performance rankings
remain similar to the baseline results.

\begin{table}[tbh]
\caption{Performance of Automatic Detection Methods in \textquotedblleft
Cleaner\textquotedblright \ Datasets}
\label{Table - Accuracy (cleaner data)}
\begin{center}
\fontsize{9pt}{11pt}\selectfont%
\begin{tabular}{lcccccccccc}
\hline \hline
& $(1)$ & $(2)$ & $(3)$ & $(4)$ & $(5)$ & $(6)$ & $(7)$ & $\left( 8\right) $
& $\left( 9\right) $ & $\left( 10\right) $ \\ 
Method & PRNR & MIMD & NMC & MBPI & FT & LS & CS & LSTM & E-RF & E-LSTM \\ 
\hline
\multicolumn{1}{c}{} &  &  &  &  &  &  &  &  &  &  \\ 
\multicolumn{11}{c}{\textit{I. Western Australia} (\# manually labeled
observations: $24,569$)} \\ 
\multicolumn{1}{c}{} &  &  &  &  &  &  &  &  &  &  \\ 
Parameter 1 & $-1.16$ & $5.14$ & $-0.20$ & $4.85$ & $0.14$ & $0.23$ & $22.50$
& $-$ & $-$ & $-$ \\ 
Parameter 2 & $-$ & $-$ & $-$ & $5$ & $-$ & $-$ & $-$ & $-$ & $-$ & $-$ \\ 
Accuracy rank & $5$ & $4$ & $8$ & $6$ & $9$ & $7$ & $10$ & $1$ & $3$ & $2$
\\ 
\% correct (median) & $90.86$ & $91.29$ & $89.44$ & $90.25$ & $88.22$ & $%
89.66$ & $85.49$ & $99.25$ & $99.06$ & $99.21$ \\ 
(Standard deviations) & $\left( 0.29\right) $ & $\left( 0.33\right) $ & $%
\left( 0.43\right) $ & $\left( 0.37\right) $ & $\left( 0.39\right) $ & $%
\left( 0.42\right) $ & $\left( 0.51\right) $ & $\left( 0.16\right) $ & $%
\left( 0.14\right) $ & $\left( 0.12\right) $ \\ 
of which cycling & $55.45$ & $56.98$ & $58.26$ & $59.69$ & $57.22$ & $56.61$
& $55.80$ & $60.09$ & $60.44$ & $60.68$ \\ 
of which not & $35.41$ & $34.31$ & $31.18$ & $30.57$ & $30.99$ & $33.05$ & $%
29.69$ & $39.15$ & $38.62$ & $38.52$ \\ 
\% false negative & $5.66$ & $3.93$ & $3.01$ & $0.53$ & $3.93$ & $3.64$ & $%
5.33$ & $0.53$ & $0.55$ & $0.49$ \\ 
\% false positive & $3.48$ & $4.78$ & $7.55$ & $9.22$ & $7.86$ & $6.70$ & $%
9.18$ & $0.22$ & $0.39$ & $0.31$ \\ 
\multicolumn{1}{c}{} &  &  &  &  &  &  &  &  &  &  \\ 
\multicolumn{11}{c}{\textit{II. New South Wales} (\# manually labeled
observations: $8,028$)} \\ 
\multicolumn{1}{c}{} &  &  &  &  &  &  &  &  &  &  \\ 
Parameter 1 & $5.24$ & $2.73$ & $1.01$ & $8.9$ & $0.22$ & $0.57$ & $4.50$ & $%
-$ & $-$ & $-$ \\ 
Parameter 2 & $-$ & $-$ & $-$ & $2$ & $-$ & $-$ & $-$ & $-$ & $-$ & $-$ \\ 
Accuracy rank & $7$ & $8$ & $10$ & $6$ & $5$ & $4$ & $9$ & $1$ & $3$ & $2$
\\ 
\% correct (median) & $91.22$ & $90.39$ & $85.65$ & $93.17$ & $93.68$ & $%
94.03$ & $88.61$ & $98.75$ & $97.44$ & $98.70$ \\ 
(Standard deviations) & $\left( 0.66\right) $ & $\left( 0.63\right) $ & $%
\left( 0.75\right) $ & $\left( 0.49\right) $ & $\left( 0.62\right) $ & $%
\left( 0.48\right) $ & $\left( 0.66\right) $ & $\left( 0.24\right) $ & $%
\left( 0.37\right) $ & $\left( 0.37\right) $ \\ 
of which cycling & $84.18$ & $84.39$ & $85.65$ & $84.18$ & $85.02$ & $83.53$
& $84.63$ & $84.56$ & $85.06$ & $85.18$ \\ 
of which not & $7.04$ & $5.99$ & $0.00$ & $8.98$ & $8.66$ & $10.50$ & $3.98$
& $14.19$ & $12.38$ & $13.53$ \\ 
\% false negative & $1.12$ & $1.19$ & $0.00$ & $1.23$ & $1.77$ & $1.95$ & $%
0.93$ & $0.75$ & $0.69$ & $0.19$ \\ 
\% false positive & $7.66$ & $8.43$ & $14.35$ & $5.60$ & $4.55$ & $4.02$ & $%
10.45$ & $0.50$ & $1.88$ & $1.11$ \\ 
\multicolumn{1}{c}{} &  &  &  &  &  &  &  &  &  &  \\ 
\multicolumn{11}{c}{\textit{III. Germany} (\# manually labeled observations: 
$22,232$)} \\ 
\multicolumn{1}{c}{} &  &  &  &  &  &  &  &  &  &  \\ 
Parameter 1 & $0.74$ & $-0.32$ & $1.26$ & $0.75$ & $0.01$ & $0.00$ & $18.50$
& $-$ & $-$ & $-$ \\ 
Parameter 2 & $-$ & $-$ & $-$ & $15$ & $-$ & $-$ & $-$ & $-$ & $-$ & $-$ \\ 
Accuracy rank & $6$ & $7$ & $10$ & $5$ & $8$ & $9$ & $4$ & $3$ & $2$ & $1$
\\ 
\% correct (median) & $68.93$ & $63.98$ & $63.44$ & $73.34$ & $63.60$ & $%
63.50$ & $81.76$ & $85.34$ & $87.41$ & $90.37$ \\ 
(Standard deviations) & $\left( 0.61\right) $ & $\left( 0.66\right) $ & $%
\left( 0.65\right) $ & $\left( 0.56\right) $ & $\left( 0.69\right) $ & $%
\left( 0.66\right) $ & $\left( 0.55\right) $ & $\left( 0.58\right) $ & $%
\left( 1.52\right) $ & $\left( 0.53\right) $ \\ 
of which cycling & $60.30$ & $62.10$ & $63.44$ & $56.42$ & $63.60$ & $63.50$
& $58.00$ & $58.09$ & $58.42$ & $59.08$ \\ 
of which not & $8.63$ & $1.88$ & $0.00$ & $16.92$ & $0.00$ & $0.00$ & $23.76$
& $27.25$ & $28.99$ & $31.28$ \\ 
\% false negative & $3.01$ & $1.48$ & $0.00$ & $6.99$ & $0.00$ & $0.00$ & $%
6.17$ & $5.39$ & $4.88$ & $3.71$ \\ 
\% false positive & $28.06$ & $34.54$ & $36.56$ & $19.66$ & $36.40$ & $36.50$
& $12.07$ & $9.27$ & $7.71$ & $5.93$ \\ \hline \hline
\end{tabular}
\begin{minipage}{475pt}
{\fontsize{9pt}{9pt}\selectfont \smallskip  \textit{Note}: These alternative results are based only on ``cleaner'' subsamples that are either unanimously labeled as ``cycling'' by all RAs or not labeled as ``cycling'' by any RA. Other details follow Table \ref{Table - Accuracy (main)}.}
\end{minipage}
\end{center}
\end{table}

\clearpage

\begin{figure}[htb!!!!]\centering%
\caption{Gains from Additional Data in ``Cleaner'' Datasets}%
\includegraphics[width=0.55\textwidth]{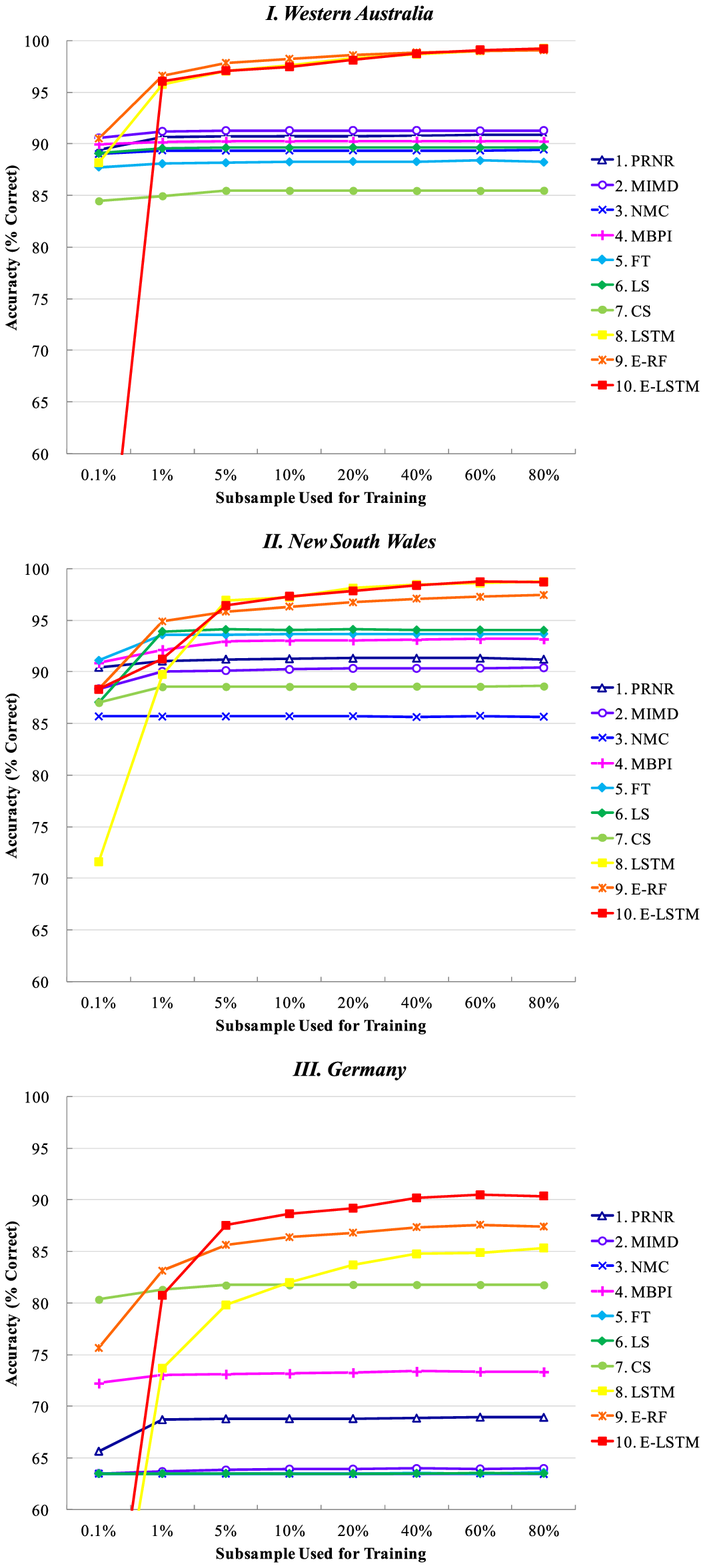}
\caption*%
{\footnotesize {\textit{Note}: These alternative results are based only on
\textquotedblleft cleaner\textquotedblright \ (i.e., less ambiguous)
subsamples that are either unanimously labeled as \textquotedblleft
cycling\textquotedblright \ by all RAs or not labeled as \textquotedblleft
cycling\textquotedblright \ by any RA.}}%
\label{Figure - Gains from Data (Cleaner Data)}
\end{figure}%

\clearpage

\section*{Appendix C \ Supplementary Plots for Section 7.2}

\paragraph*{C.1. \ Why Existing Methods Work in Australia But Fail in Germany%
}

\begin{figure}[htb!!!!]\centering%
\caption{Histograms of Median Daily Change}%
\includegraphics[width=1.00\textwidth]{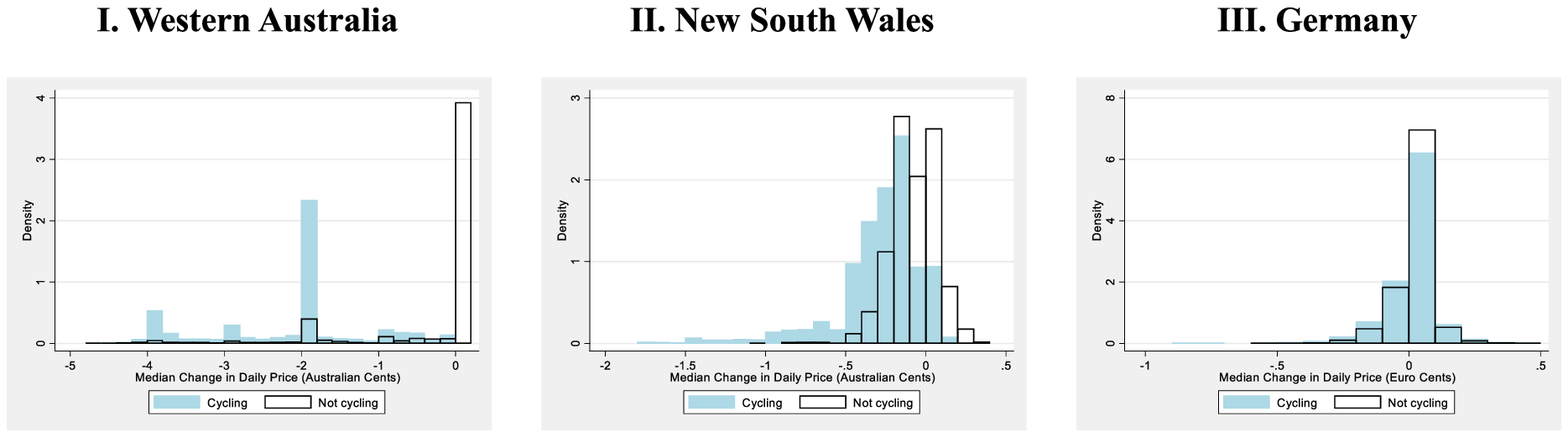}
\label{Figure - Histograms of Median Daily Change}
\end{figure}

Figure \ref{Figure - Histograms of Median Daily Change} plots histograms of
the median daily change of prices/margins, which underlies the simplest of
the asymmetry-based methods (Method 3). Most of the manually identified
cycles in Australia exhibit asymmetry (Panels I--II), whereas German cycles
are not necessarily asymmetric (Panel III).

\paragraph*{C.2. \ Why Margins Correlate Positively with Cycles in Australia
But Negatively in Germany}

The scatter plots of Figure \ref{Figure - Scatter Plots of Mean and Standard
Deviation} show the means and the standard deviations of margins are
positively correlated in all datasets. As the histograms of Figure \ref%
{Figure - Histograms of Standard Deviation} show, however, their volatility
and (manually identified) cyclicality are correlated positively in Australia
but negatively in Germany. Thus, margins and their cyclicality are
correlated positively in Australia but negatively in Germany.

\paragraph*{C.3. \ Why Do Existing Methods Find\  \textquotedblleft Positive
Correlations\textquotedblright ?}

\begin{figure}[htb!!!!]\centering%
\caption{Scatter Plots of Mean and Standard Deviation}%
\includegraphics[width=1.00\textwidth]{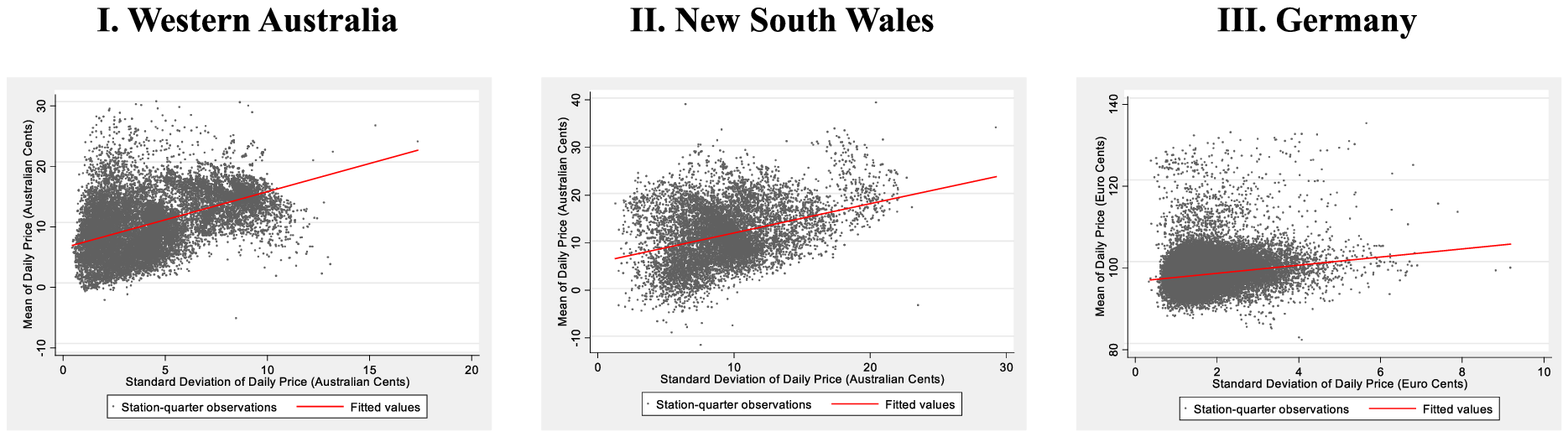}
\label{Figure - Scatter Plots of Mean and Standard Deviation}
\end{figure}

\begin{figure}[htb!!!!]\centering%
\caption{Asymmetry-based Definitions Pick Up Volatility}%
\includegraphics[width=1.00\textwidth]{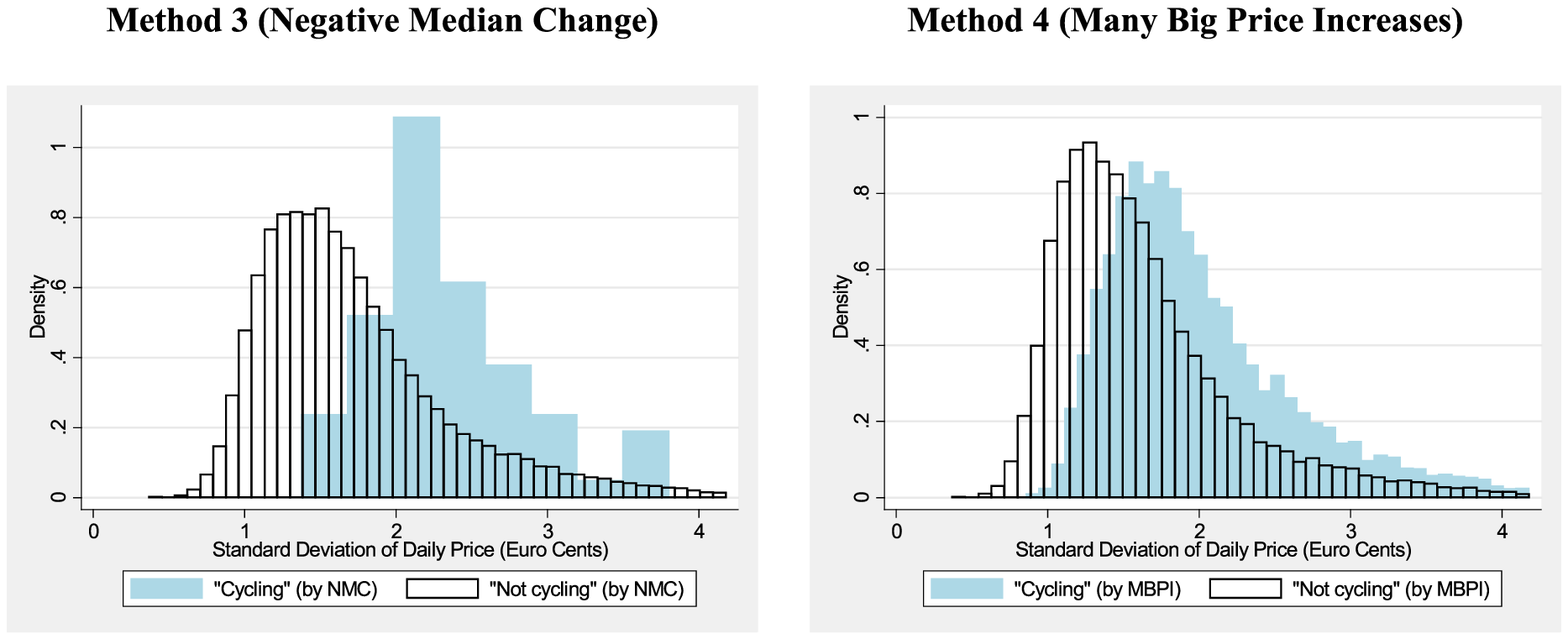}
\label{Figure - Asymmetry-based Definitions Pick Up Volatility}
\end{figure}

Figure \ref{Figure - Asymmetry-based Definitions Pick Up Volatility} shows
histograms of standard deviations by \textquotedblleft
cyclicality\textquotedblright \ based on Methods 3 and 4. These pictures
suggest the threshold rules underlying the asymmetry-based methods tend to
flag high-volatility cases as \textquotedblleft cycles,\textquotedblright \
because only sufficiently large movements can satisfy these conditions. As
Figure \ref{Figure - Scatter Plots of Mean and Standard Deviation} shows,
however, high volatility is a poor predictor of true cyclicality (based on
manual classification) in the German data.

\begin{figure}[htb!!!!]\centering%
\caption{Histograms of Standard Deviation}%
\includegraphics[width=1.00\textwidth]{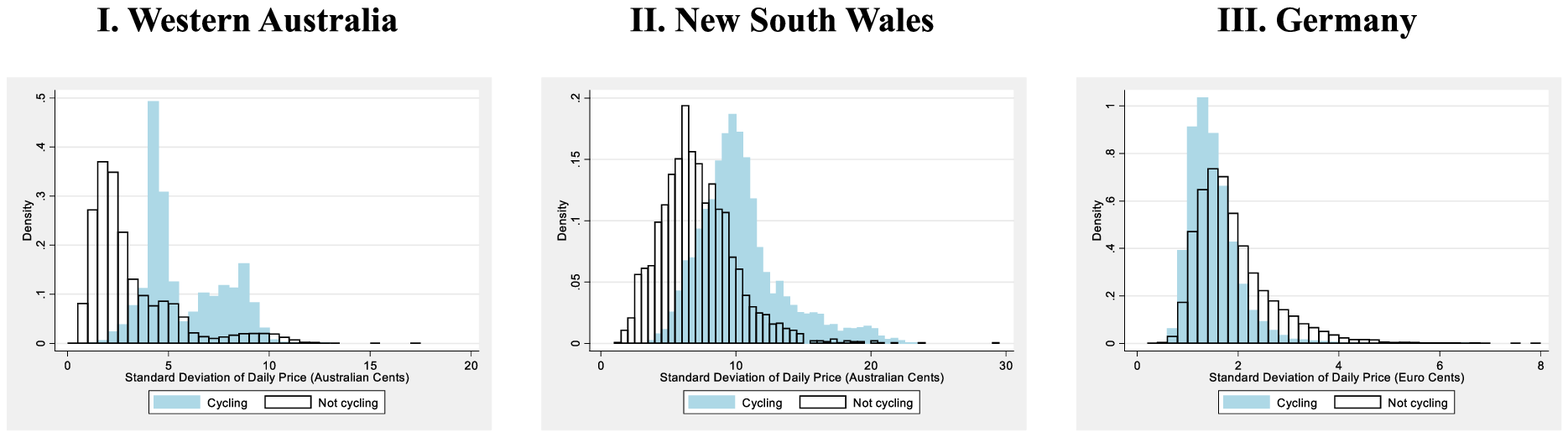}
\label{Figure - Histograms of Standard Deviation}
\end{figure}%

\end{document}